\documentclass[11pt,a4paper]{article}
\usepackage{float}
\usepackage{multirow}
\usepackage{graphics}
\usepackage[cp1252,utf8]{inputenc}
\usepackage{hyperref}
\usepackage{graphicx} %Include figure filesusepackage{graphicx} %Include figure files
\usepackage{amsmath}
\usepackage{amsfonts}
\usepackage{amssymb}
\usepackage{bigints} %big integration sign
\usepackage{relsize} %big math signs
\usepackage[top=1in, bottom=2cm, left=1.5cm, right=1.5cm]{geometry}
\usepackage{authblk}

\usepackage{changepage} %to make tables/figures start from a bit left

\usepackage{afterpage}

\usepackage{placeins}

\usepackage{wrapfig}
\title{\bf Holographic computation of Wilson loops in a background with broken conformal invariance and finite chemical potential}
\author[a]{\bf  Ashis Saha \thanks{sahaashis0007@gmail.com, ashisphys18@klyuniv.ac.in}}
\author[b]{\bf Sunandan Gangopadhyay \thanks{sunandan.gangopadhyay@gmail.com, sunandan.gangopadhyay@bose.res.in}}

\affil[a]{\textit{Department of Physics, University of Kalyani, Kalyani 741235, India}}
\affil[b]{\textit{Department of Theoretical Sciences,
 S.N.~Bose National Centre for Basic Sciences,}
\textit{JD Block, Sector-III, Salt Lake, Kolkata 700106, India}}

\date{}

\begin{document}

\maketitle

\begin{abstract}
\noindent In this paper, we follow a `bottom-up' AdS/QCD approach to holographically probe the dynamics of a moving $q\bar{q}$ pair inside a strongly coupled plasma at the boundary. We consider a deformed AdS-Reissner Nordstr\"om metric in the bulk in order to introduce nonconformality and finite quark density in the dual field theory. By boosting the gravity solution in a specific direction we consider two extreme cases of orientation, parallel and perpendicular, for the Wilson loop which in turn fixes the relative position of the $q\bar{q}$ pair with respect to the direction of boost in the plasma. By utilizing this set-up, we holographically compute the vacuum expectation value of the time-like Wilson loop in order to obtain real part of the $q\bar{q}$ potential and the effects of nonconformality (deformation parameter $c$), chemical potential $\mu$ and rapidity $\beta$ are observed on this potential. We then compute the in-medium energy loss of the moving parton (jet quenching parameter $q_m$) by setting $\beta\rightarrow\infty$ which in turn makes the Wilson loop light-like. We also use the jet quenching as an order parameter to probe the strongly-coupled domain of the dual field theory. Finally, we compute the imaginary part of the $q\bar{q}$ potential ($\mathrm{Im}(V_{q\bar{q}})$) by considering the thermal fluctuation (arbitrary long wavelength) of the string world-sheet. It is observed that for fixed values of the chemical potential and rapidity, increase in the nonconformality parameter leads to an increase in the real and imaginary potentials as well as the jet quenching parameter.
\end{abstract}

\section{Introduction}
Understanding various properties of a strongly coupled ($\lambda\equiv {g_{\mathrm{YM}}}^2N_c \gg 1$) gauge theory via AdS/CFT correspondence has been very intriguing so far. In its first avatar, the duality between a type $\mathrm{IIB}$ string theory in $\mathrm{AdS}_5\times S^5$ and $\mathcal{N}=4$ SYM in $(3+1)$-dimensions \cite{maldacena,maldacena2}, provides the tool to describe a strongly-coupled gauge theory just by choosing a proper holographic dual. Consideration of a finite temperature gauge theory holographically demands the presence of a black hole in the gravity theory. It also states that if the gauge theory is strongly coupled then the supergravity (holographic dual spacetime geometry) will be weakly coupled or vice-versa \cite{witten}.\\
On the other hand it is now a well accepted fact that in the relativistic heavy ion collider (RHIC), a new state of matter known as quark-gluon plasma (QGP) is formed at the center of the collision \cite{qgp}-\cite{qgp3}. The high temperature and high density phases of QGP are dominated by quarks and gluons. It is observed that the binding interaction of a heavy quark-antiquark ($q\bar{q}$) pair is screened by the QGP which leads to the dissociation of the heavy quarkonium \cite{satz}. It is also realized that one has to consider the motion of the medium during the production of the $q\bar{q}$ pair in the strongly coupled plasma together with the the in-medium interaction. Observations of various paramters, namely, the study of in-medium loss of the moving partons (jet quenching), melting of quarkonium, thermalization, and so on, has been playing key roles as observables in the RHIC experiments and QCD. It has been observed that the nature of QGP is strongly coupled. The only methods to study a strongly coupled phenomena in QCD are lattice simulations and effective field theories. A completely new approach to study strongly coupled QCD is AdS/CFT conjecture due to the strong/weak dual mapping. In particular, the study of the binding energy of a $q\bar{q}$ (quark-antiquark) pair, screening length, energy loss of a moving parton in the high $p_T$ limit (jet quenching) by utilizing AdS/CFT conjecture has been a matter of growing interest in recent times. Along with these observables, the computation of imaginary potential via gauge/gravity is also a matter of great importance from QCD phenomenlogical point of view as it leads to the computation of the thermal width in QCD. The thermal width plays an important role in the study of decay processes. The original calculation of $q\bar{q}$ binding energy in a supersymmetric $\mathcal{N}=4$ Yang-Mills theory via expectation value of a time-like Wilson loop was first done in \cite{maldacena3} which was later extended to the finite temperature case in \cite{finite1}-\cite{finite3}. The holographic computation of the screening length of the moving $q\bar{q}$ inside a moving hot plasma was done in \cite{moving1,moving2}. The observable of in-medium energy loss of a moving parton or the jet quenching parameter was computed in \cite{liu1,liu2}  where it was shown that the jet quenching is related to the light-like Wilson loop. The holographic study of jet quenching parameter has been generalized to various dual geometries \cite{ex1}-\cite{ex8} in accordance to the property of the boundary gauge theory. In \cite{eq}, it was shown that the jet quenching parameter can be treated as an order parameter to decide whether a theory is strongly-coupled or weakly-coupled. The scenario has also been extended to introduction of charged black holes in the dual geometry in order to include finite quark density in the gauge theory. The chemical potential $\mu$ appears as the quark density operator and according to the AdS/CFT conjecture, a vector operator in the boundary field theory is dual to a gauge field in the bulk. This motivates to include a $U(1)$ gauge field in the Einstein-Hilbert action of the gravity which leads to the well-known solution AdS Reissner Nordstrom black hole \cite{c1}-\cite{c3}.\\
The computation of imaginary potential via thermal fluctuation of the string world-sheet was first shown in \cite{im1} in which it was also shown that imaginary part of the $q\bar{q}$ potential leads to the thermal width by considering a Coulombic wave-function. Some important studies on imaginary $q\bar{q}$ potential can be found in \cite{im2}-\cite{im7}. For instance, in \cite{movingc}, the effect of chemical potential on the imaginary part of the moving $q\bar{q}$ pair potential has been observed. It is found that at fixed chemical potential, increasing the rapidity decreases the value of the imaginary potential. Also at fixed value of the rapidity parameter, increasing the chemical potential reduces the value of the imaginary potential. It has also been pointed out that the effects are more prominent in case of the perpendicular case compared to the parallel case.\\
However, if a theory is nonconformal, the choice of a geometry which shall be dual to a strongly coupled QCD is somewhat tricky. In this paper we follow a `bottom-up' approach of AdS/QCD correspondence. We use the soft-wall ($\mathrm{SW}_{T,\mu}$) geometry in which the overall metric is multiplied with warp factor (quadratic dilaton) in order to probe more realistic picture of QCD. These type of models \cite{hqcd1}-\cite{hqcd4} were introduced in order to emulate confinement at zero temperature and have achieved considerable success in describing various aspects of hadron physics. The $\mathrm{SW}_{T,\mu}$ model mentioned in this paper, also introduces finite quark density in the dual field theory due to the presence of chemical potential $\mu$. The study of the jet quenching in a nonconformal medium was carried out in \cite{buchel}. In \cite{hqcd5}-\cite{hqcd4}, the confinement/deconfinement phase transition and QCD phase diagram has been obtained using the $\mathrm{SW}_{T,\mu}$ dual geometry set up. Studies in this direction can further be found in \cite{mu1,mu2}. The calculation of the imaginary potential in this dual geometry for heavy quark was carried out in \cite{hqcd5}. In this paper, we find that for a fixed value of nonconformality parameter, the effect of increasing the  chemical potential with a fixed value of rapidity decreases the value of the real potential, decreases the imaginary potential but increases the in-medium energy loss. These observations are in conformity with the previous findings \cite{realq,mu3,movingc,hqcd5}. Further, for fixed values of the chemical potential and rapidity, increasing the nonconformality parameter results in the increase in the real potential, imaginary potential and the jet quenching parameter.\\
The paper is organized as follows. In section(\ref{sec1}) we breifly discuss the dual geometry ($\mathrm{SW}_{T,\mu}$) and introduce Lorentz boost to the metric in order probe a moving $q\bar{q}$ pair. Furthermore, we choose two extreme cases for the orientation of the $q\bar{q}$ pair, parallel to the direction of boost and perpendicular to the direction of boost. In section (\ref{20}), we compute the screening length and real $q\bar{q}$ potential for both parallel and perpendicular case. We represent the results graphically. In section(\ref{sec3}), we compute the jet quenching parameter for $\mathrm{SW}_{T,\mu}$ model and study the effects of chemical potential, nonconformality on it. The computation of the imaginary potential is done in section(\ref{sec4}) and the results are represented graphically. Finally, we conclude in section(\ref{sec6}). We also have an appendix in the paper.

\section{Soft-wall dual geometry model}\label{sec1}
We start from a geometry in the bulk in order to represent the basic characteristics of the boundary $\mathrm{SU}(N)$ gauge theory in the large $N$ limit. The motivation of starting with this bulk spacetime is to study the dynamics of a moving $q\bar{q}$ pair in a nonconformal, strongly coupled Yang-Mills theory with finite quark density. The study of such a strongly coupled Yang-Mills theory, namely QCD, is in general a difficult problem. However, the gauge/gravity correspondence, also known as the AdS/CFT duality, tries to provide a solution to this difficult problem by means of a gravitational dual in a higher dimensional spacetime. It should be mentioned however, that a geometric dual to QCD is not known and these gravitational duals with which one works are toy models that reproduces the phenomenological aspects of the strongly coupled QCD. For example, since QCD is nonconformal, the dual gravitational model should also have broken conformal invariance. Moreover, to get linear confinement at zero temperature and have a mass scale related to $\Lambda_{\mathrm{QCD}}$, a five-dimensional gravitational theory on AdS$_5$ spacetime is used with a warp factor. Further, it is known from QCD that the term $J_{D}=\mu \psi^{+}(x)\psi(x)$ has to be added to the Lagrangian in order to incorporate the effect of finite quark density. Here $\mu$ represents the chemical potential and acts as the source of the quark density operator. The AdS/CFT dictionary says that the source of an operator is dual to the non-normalizable mode of a dual field in the bulk. Since we shall consider the dual field in the bulk to be massless, the source of a QCD operator is the boundary value of this massless dual field in the bulk. Hence, the chemical potential can be considered as the boundary value of the time component of a $U(1)$ gauge field $A_{\mu}~ (\mu=0,1,...,4)$. This gauge field in the bulk is dual to the vector quark current.\\
Before introducing the $\mathrm{SW}_{T,\mu}$ model, to fix up the notations we begin by writing down the action of holographic models dual to $3+1$-dimensional boundary field theory with chemical potential $\mu$. This reads
 \begin{eqnarray}
 S_{Bulk} = \frac{1}{16\pi G_5} \int d^5x \sqrt{-g}(\mathcal{R}-2\Lambda-\frac{1}{4}\mathrm{F}_{\mu\nu}\mathrm{F}^{\mu\nu})~.
 \end{eqnarray}
The above action leads to the following equation of motion 
 \begin{eqnarray}\label{eom}
 R_{\mu\nu} -\frac{1}{2}(\mathcal{R}-2\Lambda) g_{\mu\nu}&=&g^{\alpha\beta}F_{\alpha\mu}F_{\beta\nu}-\frac{1}{4}g_{\mu\nu}(F^{\rho\eta}F_{\rho\eta})
 \end{eqnarray}
 \begin{eqnarray}
 \partial_{\alpha} \big(\sqrt{-g}g^{\mu\alpha}g^{\nu\beta}F_{\alpha\beta}\big)&=&0
 \end{eqnarray}
where
\begin{eqnarray}
F_{\mu\nu}= \partial_{\mu} A_{\nu} - \partial_{\nu} A_{\mu}
\end{eqnarray}
is the Maxwell field strength tensor and $A_{\mu}$ represents the $U(1)$ gauge field in the bulk. The solution of the equation of motion given in eq.(\ref{eom}) leads to the $AdS_{4+1}$ Reissner-Nordstr\"om spacetime, which can be written down in the planar coordinates as
\begin{eqnarray}\label{metric}
ds^2 = \frac{r^2}{R^2}\big[-f(r)dt^2+d\vec{x}^2\big] + \frac{R^2}{r^2f(r)}dr^2
\end{eqnarray}
where $f(r)$ is given by
\begin{eqnarray}
f(r)= 1-(1+Q^2) (\frac{r_h}{r})^4 +Q^2 (\frac{r_h}{r})^6
\end{eqnarray}
with $r_h$ representing the event horizon of the black hole and $Q= \frac{q}{r_h^3}$. The choice for the U(1) gauge field reads
\begin{eqnarray}
A= \left(-k\frac{q}{r^2}+\Phi\right)dt~.
\end{eqnarray}
where $k$ is a dimensionaless parameter. The fact that the gauge field vanishes at $r=r_h$ yields
	\begin{eqnarray}
	\Phi=k\frac{q}{r_h^2}~.
	\end{eqnarray} 
	From the AdS/CFT dictionary, the boundary value of the time component of the $U(1)$ gauge field $A_{\mu}$ is considered to be the chemical potential in the boundary theory. Hence we have
\begin{eqnarray}\label{eq9}
	\mu \equiv \lim_{r \to \infty} A_t = \frac{kq}{r_h^2} = k Q r_h~.
\end{eqnarray} 
The time component of the gauge field $A_{\mu}$ can now be expressed in terms of the chemical potential as
\begin{eqnarray}
A_t = \mu \bigg(1-\frac{r_h^2}{r^2}\bigg)~.
\end{eqnarray}
\noindent Eq.(\ref{eq9}) provides the following relationship between the charge of the black hole $Q$ and the chemical potential $\mu$
\begin{eqnarray}\label{cmu}
Q= \frac{\mu }{\sqrt{3} r_h}
\end{eqnarray}
where we set the dimensionless parameter $k=\sqrt{3}$ \cite{movingc}, also we have set the AdS radius $R=1$.
Eq.(\ref{cmu}) relates the charge of the black hole $Q$ to the chemical potential $\mu$ which is measure of the finite quark density in the dual field theory. By using eq.(\ref{cmu}), we can write the Hawking temperature of the black hole in terms of the chemical potential $\mu$ as
\begin{eqnarray}\label{ht}
	T_H= \frac{r_h}{\pi}\left(1-\frac{\mu^2}{6r_h^2}\right)~.
\end{eqnarray}
With the help of eq.(\ref{ht}) we can express the event horizon radius $r_h$ in terms of the chemical potential $\mu$ and Hawking temperature $T_H$ as
\begin{eqnarray}\label{eh}
r_h = T_H \bigg[\frac{\pi}{2}+ \sqrt{\frac{\pi^2}{4}+\frac{1}{6}\left(\frac{\mu}{T_H}\right)^2}\bigg]~.
\end{eqnarray}
The effect of confinement in the boundary theory can be emulated by introducing a warp factor $h(r)$ to the black hole spacetime \eqref{metric}. The introduction of this factor in this spacetime also breaks the conformal invariance in the boundary field theory. This leads to a class of two-parameter deformed soft-wall models ($\mathrm{SW}_{T,\mu}$). These models include the effects of finite quark density. The metric of these model reads \cite{hqcd1,hqcd4}
\begin{eqnarray}\label{metric1}
ds^2 = \frac{r^2 h(r)}{R^2}\big[-f(r)dt^2+d\vec{x}^2\big] + \frac{R^2h(r)}{r^2f(r)}dr^2~;~~~ h(r) = \exp\left(\frac{C^2}{r^2}R^4\right)
\end{eqnarray}  
where $C$ represents the deformation parameter having the dimension of energy. As there is only one deformation parameter to incorporate nonconformality, we shall call it a class of one-parameter $\mathrm{SW}_{T, \mu}$ models. The knowledge of lattice field theory suggests that the range $0\leq \frac{C}{T} \leq 2.5$ is the relevant region to compare $\mathrm{SW}_{T,\mu}$ models with QCD \cite{robust}. We shall use this relevant region of $\frac{C}{T}$ through out this work. It is to be noted that in the limit $C \rightarrow 0$, we obtain the usual Reissner-Nordstr\"om metric in the bulk.\\
The scenario of a moving $q\bar{q}$ pair can be understood from a frame in which the plasma is at rest and the dipole $q\bar{q}$ is moving. Equivalently, we can boost to a reference frame in which the $q\bar{q}$ dipole is at rest but the finite temperature plasma is moving. In this set up, a Lorentz boost to the reference frame in the $x_1$ direction with the rapidity $\beta$ which can be expressed as
\begin{eqnarray}
	dt &=& \cosh{\beta} ~dt^{\prime} - \sinh\beta~ dx_1^{\prime}\nonumber\\
	dx_1 &=& -\sinh{\beta}~dt^{\prime} + \cosh{\beta}~dx_1^{\prime}\nonumber~.
\end{eqnarray} 
The $q\bar{q}$ is at rest and feels a hot plasma wind as the plasma is moving with a velocity $v=\tanh\beta$ in the $-x_1^{\prime}$ direction. The metric given in eq.(\ref{metric1}) reads in the boosted coordinates ($t^{\prime}, x_1^{\prime}$) reads
\begin{eqnarray}\label{boosted}
ds^2 &=& -r^2 h(r) \big[f(r)\cosh^2{\beta} - \sinh^2{\beta}\big]dt^2 + r^2 h(r)\big[\cosh^2{\beta} -f(r) \sinh^2{\beta}\big]dx_1^2\nonumber\\
&& + r^2 h(r) \big[dx_2^2+dx_3^2\big] + \frac{h(r)}{r^2 f(r)} dr^2 - 2r^2 h(r) \cosh{\beta}\sinh\beta\big[1-f(r)\big]dt dx_1
\end{eqnarray}
where we set $AdS$ radius $R=1$ and dropped the primes ($\prime$) in the coordinates for the sake of simplicity. Keeping in mind the presence of boost, we choose two the following $q\bar{q}$ pair orientations, namely, 
\begin{eqnarray}
&&q\bar{q}~\mathrm{dipole~is~parallel~to~the~direction ~of ~boost}\nonumber\\
&&q\bar{q}~\mathrm{dipole~is~perpendicular~to~the~direction ~of ~boost}\nonumber~.
\end{eqnarray}
In the following section we use this geometry to compute the screening length and the real part of the $q\bar{q}$ potential in a nonconformal plasma with finite density properties.
\section{Screening length and the real part of the $q\bar{q}$ potential}\label{20}
In this section, we  set out to evaluate the real part of the $q\bar{q}$ potential holographically. The field theoretic approach to obtain this is to evaluate the expectation value of the Wilson loop operator 
\begin{eqnarray}
W[\mathcal{C}]=\frac{1}{N}\mathrm{Tr}~\mathrm{P}e^{i\oint_\mathcal{C} A_{\mu}dx^{\mu}}
\end{eqnarray} 
where $\mathcal{C}$ is a rectangular loop in spacetime and the trace is over the fundamental representation of the SU(N) group \cite{maldacena3}. In the limit $\tau\to\infty$ this expectation value is given by
\begin{eqnarray}
\langle W[\mathcal{C}]\rangle = e^{-iV_{q\bar{q}}(L)\mathcal{T}}
\end{eqnarray}
where $V_{q\bar{q}}(L)$ represents the real part of the $q\bar{q}$ potential.\\
\noindent In the holographic prescription, the expectation value of the rectangular Wilson loop is computed by probing a string in the gravity background. It is assumed that the string moves in the direction $t$ with a velocity $v$ ($0 < v< 1$) with the endpoints of the strings lies in the $x_{n}$ direction ($n$ depends upon the choice of the `orientation' of the $q\bar{q}$ pair with respect to the direction of the Lorentz boost) with a spatial separation $L$. The end points of this open string represents the quark-antiquark pair in the plasma. The length of this rectangular Wilson loop is specified as $\mathcal{T}$ and $L$ in the directions $t$ and $x_n$ respectively. Furthermore, it is to be noted that $L\ll\mathcal{T}$ so that the string worldsheet has time translational invariance. With this rectangular contour $\mathcal{C}$ ($\mathcal{T}-L$) in place, the Nambu-Goto action of the fundamental probe, which in this case is the open string, is computed. This is related to the expectation value of the Wilson loop (in the supergravity limit) as \cite{maldacena3},\cite{finite1}
\begin{eqnarray}
\langle W[\mathcal{C}]\rangle = e^{-iS_{I}}
\end{eqnarray}
where $S_I$ is the regularized Nambu-Goto action of the open string.\\
\subsection{$q\bar{q}$ pair is parallel to the boost direction}
To begin our study, we first choose $x_n$ to be $x_1$. This choice phyically represents the fact that the $q\bar{q}$ pair is oriented along the direction of the boost which in turn also specifies that the Wilson loop lies in the $t-x_1$ plane.\\
 The Nambu-Goto string world-sheet action reads
\begin{eqnarray}\label{action}
S_{NG} = \frac{1}{2\pi \alpha^{\prime}} \int \sqrt{-\mathrm{det}g_{ab}}~ d\sigma d\tau
\end{eqnarray} 
where $g_{ab}$ is the induced metric on the target space given by
\begin{eqnarray}
g_{ab}= G_{\rho\eta} \frac{\partial x^{\rho}}{\partial\xi^{a}}\frac{\partial x^{\eta}}{\partial\xi^{b}};~~x^{\rho}(\sigma,\tau)\equiv\mathrm{Worldsheet~embedding~coordinates}
\end{eqnarray}
$G_{\rho\eta}$ in the above relation represents the metric given in eq.(\ref{boosted}). The static gauge $\sigma=x_1$, $\tau=t$, $r=r(\sigma)$, $x_2=x_3=\mathrm{fixed}$ is chosen to represent the parallel orientation of the $q\bar{q}$ pair. We also require boundary conditions. The open string with profile $r=r(\sigma)$ has the following boundary conditions
\begin{eqnarray}\label{bc}
r(\sigma\equiv x_1 = \pm \frac{L}{2}) = \infty~.
\end{eqnarray}
With the above set up in place, we compute the Nambu-Goto action \ref{action}. This is given by
\begin{eqnarray}\label{ng1}
S_{NG} = \frac{\mathcal{T}}{2\pi\alpha^{\prime}}\int_{-L/2}^{+L/2} d\sigma\mathcal{L}(r,r^{\prime}) = \frac{\mathcal{T}}{\pi\alpha^{\prime}}\int_{0}^{L/2} d\sigma\sqrt{A(r)+B(r)~{r^{\prime}}^2};~r^{\prime}\equiv \frac{dr}{d\sigma}
\end{eqnarray}
where
\begin{eqnarray}\label{functions}
A(r) &=& r^4 h^2(r) f(r) \nonumber\\
B(r) &=& \frac{h^2(r)}{f(r)}\big(f(r)\cosh^2\beta-\sinh^2\beta\big)~.
\end{eqnarray}
The Lagrangian $\mathcal{L}(r,r^{\prime})$ leads to the following Hamiltonian obtained by a Legendre transformation
\begin{eqnarray}
\mathcal{L} - r^{\prime} \frac{\partial \mathcal{L}}{\partial r^{\prime}} = \frac{A(r)}{\sqrt{A(r) +B(r) {r^{\prime}}^2}} \equiv \mathcal{H} = \mathrm{constant}~.
\end{eqnarray}
Now the condition for the turning point of the open string, which is the furthest part of the U-shaped string profile reads
\begin{eqnarray}
\frac{dr}{d\sigma} =0~\mathrm{at}~r=r_t
\end{eqnarray}
where $r=r_t$ is the turning point of the $\mathrm{U}$-shaped string profile inside the bulk. This fixes the constant to be
\begin{eqnarray}
\mathcal{H} = \sqrt{A(r_t)}~;~~A(r_t) \equiv A(r=r_t)~.
\end{eqnarray}
This leads to
\begin{eqnarray}\label{1}
\frac{dr}{d\sigma} &=& \sqrt{\frac{A(r)}{B(r)}} \sqrt{\bigg[\frac{A(r)}{A(r_t)}-1\bigg]}~~\nonumber\\
&=& \frac{r^2f(r)}{\sqrt{f(r)\cosh^2\beta-\sinh^2\beta}}\sqrt{\bigg[\frac{r^4 h^2(r) f(r)}{r_t^4 h^2(r_t) f(r_t)}-1\bigg]}
\end{eqnarray}
where in the second line of the equality we have used eq.(\ref{functions}).
Integrating eq.(\ref{1}) and using the boundary condition \ref{bc}, we obtain the screening length $L$ as
\begin{eqnarray}\label{lpara}
L = 2 r_t^2 h(r_t) \sqrt{f(r_t)}\int_{r_t}^{\infty} \frac{\sqrt{f(r)\cosh^2\beta-\sinh^2\beta}}{r^4 f(r) h(r) \sqrt{f(r) - (\frac{r_t}{r})^4(\frac{h(r_t)}{h(r)})^2 f(r_t)}}~dr~.
\end{eqnarray}
It is important to note that $\frac{dr}{d\sigma}$ encounters a singularity at $r=r_c$ given by
\begin{eqnarray}
f(r_c)\cosh^2\beta-\sinh^2\beta = 0~.
\end{eqnarray}
The presence of this singular point $r_c$ restricts the domain of values for the turning point $r_t$ since $r_t<r_c$ will lead to a problem in obtaining the screening length $L$. So the turning point $r_t$ must satisfy the condition $r_t>r_c$ so that we get a physically relevant value for the screening length $L$ \cite{ex2}. This condition for the turning point $r_t$ further imposes the condition (since $f(r)$ is a monotonically increasing function of $r$)
\begin{eqnarray}\label{tpc}
f(r_t)\cosh^2\beta-\sinh^2\beta &>& 0\\ \nonumber~.
f(r_t)&>&\tanh^2{\beta}~.
\end{eqnarray}
This constraint will be taken into account throughout the subsequent discussion.
\noindent Defining $\frac{r_t}{r} = u$ and $\frac{r_h}{r_t} = \alpha$ eq.(\ref{lpara}) takes the form
\begin{eqnarray}\label{L1}
LT=2\big(\frac{\alpha}{t_{\mu}}\big) \sqrt{f(\alpha)} \int_{0}^{1} du \frac{u^2 \exp[(\frac{\alpha}{t_{\mu}})^2 (\frac{C}{T})^2 (1-u^2)]\sqrt{f(u)\cosh^2\beta-\sinh^2\beta}}{f(u)\sqrt{f(u)-u^4 f(\alpha) \exp[2(\frac{\alpha}{t_{\mu}})^2 (\frac{C}{T})^2 (1-u^2)]}}
\end{eqnarray}
where
\begin{eqnarray}
r_h&=&T \bigg[\frac{\pi}{2}+ \sqrt{\frac{\pi^2}{4}+\frac{1}{6}(\frac{\mu}{T})^2}\bigg] \equiv T~ t_{\mu}\nonumber\\
f(u)&=& 1-\bigg[1+\frac{1}{3{t_{\mu}}^2}\bigg(\frac{\mu}{T}\bigg)^2\bigg]\alpha^4 u^4 + \frac{1}{3{t_{\mu}}^2}\bigg(\frac{\mu}{T}\bigg)^2 \alpha^6 u^6\nonumber\\
f(\alpha) &=& 1- \bigg[1+\frac{1}{3{t_{\mu}}^2}\bigg(\frac{\mu}{T}\bigg)^2\bigg]\alpha^4+\frac{1}{3{t_{\mu}}^2}\bigg(\frac{\mu}{T}\bigg)^2\alpha^6~.
\end{eqnarray}
Note that the $\mathrm{gauge/gravity}$ duality dictionary is used to denote the Hawking temperature of the black hole $T_H$ as the temperature $T$ of the plasma. 
\noindent Now substituting $r^{\prime}=\frac{dr}{d\sigma}$ from eq.(\ref{1}) in the Nambu-Goto action \eqref{ng1}, we obtain
\begin{eqnarray}
S_{NG} = \frac{\mathcal{T}}{\pi\alpha^{\prime}}	\int_{0}^{L/2} dr \frac{h(r) \sqrt{f(r)\cosh^2\beta-\sinh^2\beta}}{\sqrt{f(r) - (\frac{r_t}{r})^4(\frac{h(r_t)}{h(r)})^2 f(r_t)}}~.
\end{eqnarray}
Once again setting $\frac{r_t}{r} = u$ and $\frac{r_h}{r_t} = \alpha$, the string world-sheet action takes the form
\begin{eqnarray}\label{UnregE}
S_{NG} = \frac{\mathcal{T}}{\pi \alpha}\sqrt{\lambda}~T t_{\mu}\int_{0}^{1} du \frac{ \exp[(\frac{\alpha}{t_{\mu}})^2 (\frac{C}{T})^2 (1-u^2)]\sqrt{f(u)\cosh^2\beta-\sinh^2\beta}}{u^2\sqrt{f(u)-u^4 f(\alpha) \exp[2(\frac{\alpha}{t_{\mu}})^2 (\frac{C}{T})^2 (1-u^2)]}}~.
\end{eqnarray}
In writing down the above expression, we have used the gauge/gravity dictionary to relate the t'Hooft coupling constant with the string tension as
\begin{eqnarray}
\sqrt{\lambda} = \frac{1}{\alpha^{\prime}}~.
\end{eqnarray}
The above string action is divergent as it contains the self-energy contribution from the $q\bar{q}$ pair. In order to regularize it, we need to substract the self energies of the two individual quarks. The self energy term $S_0$ reads
\begin{eqnarray}\label{selfE}
S_0 = \frac{\tau}{\pi \alpha^{\prime}} \int_{r_h}^{\infty} \sqrt{-g_{tt}g_{rr}}\vert_{r\rightarrow\infty} ~dr
\end{eqnarray}
where the metric components $g_{tt}$ and $g_{rr}$ are given by
\begin{eqnarray}
g_{tt} &=& - r^2 h(r) \big[f(r)\cosh^2{\beta} - \sinh^2{\beta}\big] \nonumber\\
g_{rr} &=& \frac{h(r)}{r^2 f(r)}~.
\end{eqnarray}
\noindent Substracting eq.(\ref{selfE}) from eq.(\ref*{UnregE}), we obtain the regularized open string action reads\\
\begin{eqnarray}
S_I = \frac{\mathcal{T}}{\pi \alpha}\sqrt{\lambda}~T t_{\mu}\bigg(\int_{0}^{1} \frac{1}{u^2}\bigg[\frac{ \exp[(\frac{\alpha}{t_{\mu}})^2 (\frac{C}{T})^2 (1-u^2)]\sqrt{f(u)\cosh^2\beta-\sinh^2\beta}}{\sqrt{f(u)-u^4 f(\alpha) \exp[2(\frac{\alpha}{t_{\mu}})^2 (\frac{C}{T})^2 (1-u^2)]}}-1\bigg]du-1+\alpha\bigg)~.
\end{eqnarray}
\noindent This leads to the real part of the $q\bar{q}$ pair potential to be
\begin{eqnarray}\label{ReP1}
\frac{\mathrm{Re}\big( V_{q\breve{q}}\big)}{\sqrt{\lambda}T} =\bigg(\frac{t_{\mu}}{\alpha}\bigg) \bigg(\int_{0}^{1} \frac{1}{u^2}\bigg[\frac{ \exp[(\frac{\alpha}{t_{\mu}})^2 (\frac{C}{T})^2 (1-u^2)]\sqrt{f(u)\cosh^2\beta-\sinh^2\beta}}{\sqrt{f(u)-u^4 f(\alpha) \exp[2(\frac{\alpha}{t_{\mu}})^2 (\frac{C}{T})^2 (1-u^2)]}}-1\bigg]du-1+\alpha\bigg)~.
\end{eqnarray}
The above expression represents the effects of the chemical potential $\mu$, rapidity $\beta$ and nonconformality $\frac{C}{T}$ on the real part of the $q\bar{q}$ pair potential moving along the direction of the Lorentz boost.\\
 We shall now compute the same for the thing in the scenario representing the moving $q\bar{q}$ pair in the direction perpendicular to the Lorentz boost. 
\subsection{$q\bar{q}$ pair is perpendicular (transverse) to the boost direction}
In this case, we choose the following parametrizations for the string action
\begin{eqnarray}
\sigma= x_3,~~\tau= t,~~r=r(\sigma)~ \mathrm{and}~ x_2 = x_1 =\mathrm{constant}\nonumber
\end{eqnarray}
The open string action \ref{action} now reads
\begin{eqnarray}\label{SNG}
S_{NG} = \frac{\mathcal{T}}{2\pi\alpha^{\prime}}\int_{-L/2}^{+L/2} d\sigma\mathcal{L}(r,r^{\prime}) \equiv \frac{\mathcal{T}}{\pi\alpha^{\prime}}\int_{0}^{L/2} dx_3\sqrt{A(r)+B(r)~{r^{\prime}}^2}
\end{eqnarray}
where
\begin{eqnarray}
A(r) &=& r^4 h^2(r) \big(f(r)\cosh^2\beta - \sinh^2\beta\big)\nonumber\\
B(r) &=& \frac{h^2(r)}{f(r)}\big(f(r)\cosh^2\beta - \sinh^2\beta\big)~. 
\end{eqnarray}
Following the procedure in the earlier section, we get 
\begin{eqnarray}
\frac{dr}{d\sigma} = \sqrt{\frac{A(r)}{B(r)}}\sqrt{\left(\frac{h(r)}{h(r_t)}\right)^2 \frac{A(r)}{A(r_t)}-1}~.
\end{eqnarray}
The screening length $LT$ is now given by
\begin{eqnarray}\label{L2}
LT&=&2\big(\frac{\alpha}{t_{\mu}}\big) \sqrt{f(\alpha)\cosh^2\beta - \sinh^2\beta}\nonumber\\
&&\int_{0}^{1} du \frac{u^2 \exp[(\frac{\alpha}{t_{\mu}})^2 (\frac{C}{T})^2 (1-u^2)]}{\sqrt{f(u)}\sqrt{f(u)\cosh^2\beta - \sinh^2\beta-u^4 (f(\alpha)\cosh^2\beta - \sinh^2\beta) \exp[2(\frac{\alpha}{t_{\mu}})^2 (\frac{C}{T})^2 (1-u^2)]}}~.
\end{eqnarray}
The real part of the $q\bar{q}$ pair potential therefore reads
\begin{eqnarray}\label{ReP2}
\frac{\mathrm{Re}\big( V_{q\breve{q}}\big)}{\sqrt{\lambda}T} &=&\bigg(\frac{t_{\mu}}{\alpha}\bigg) \times \nonumber\\
&&\bigg(\int_{0}^{1} \frac{du}{u^2}\bigg[\frac{ \exp[(\frac{\alpha}{t_{\mu}})^2 (\frac{C}{T})^2 (1-u^2)](f(u)\cosh^2\beta-\sinh^2\beta)}{\sqrt{f(u)}\sqrt{f(u)\cosh^2\beta-\sinh^2\beta-u^4 (f(\alpha)\cosh^2\beta-\sinh^2\beta) \exp[2(\frac{\alpha}{t_{\mu}})^2 (\frac{C}{T})^2 (1-u^2)]}}\nonumber\\
&&-1\bigg]-1+\alpha\bigg)~.
\end{eqnarray} 
The expressions for the $q\bar{q}$ separation length $L$ (eq.(s)(\ref{L1}, \ref{L2})), and also the expressions for the real $q\bar{q}$ pair potential (eq.(s)(\ref{ReP1}, \ref{ReP2})) cannot be computed analytically. Hence, we take recourse to numerical study to see the effects of rapidity $\beta$, nonconformality $\frac{C}{T}$ and chemical potential $\mu$. We do this in the next section. Interestingly, we observe that in the perpendicular case there is no constraint on the turning point $r_t$ unlike the parallel case.

\subsection{Numerical analysis and observations}
In this subsection, we shall carry out numerical analysis in order to draw proper conclusions. We have cast the expressions for the screening lengths and potentials for both perpendicular and parallel cases in a dimensionless form in order to avoid any ambiguity in the analysis. In the dimensionless form, the deformation parameter $C$ which have the dimension of energy, now becomes $\frac{C}{T}$. The same is true for the chemical potential $\mu$ which now becomes $\frac{\mu}{T}$.
\begin{figure}[!h]
	\begin{minipage}[t]{0.48\textwidth}
		\centering\includegraphics[width=\textwidth]{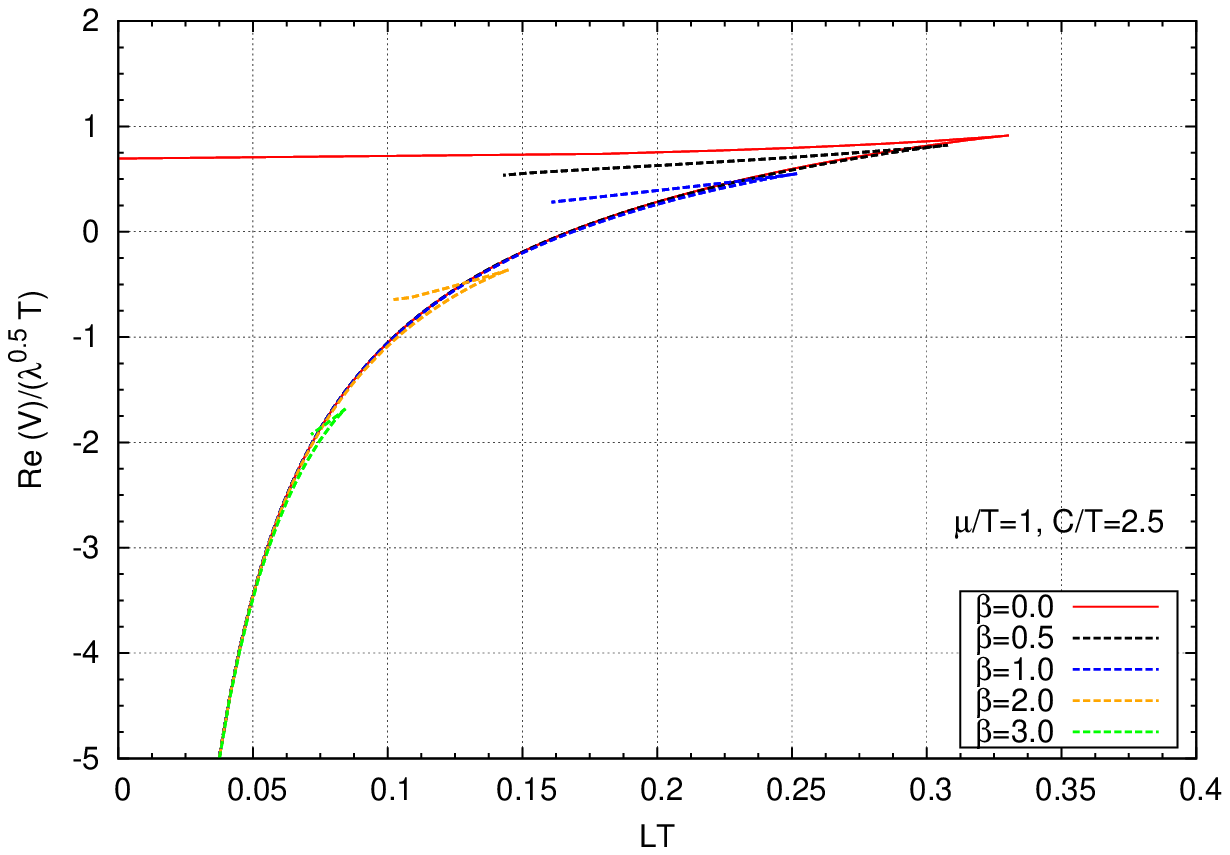}\\
		{\footnotesize parallel case}
	\end{minipage}\hfill
	\begin{minipage}[t]{0.48\textwidth}
		\centering\includegraphics[width=\textwidth]{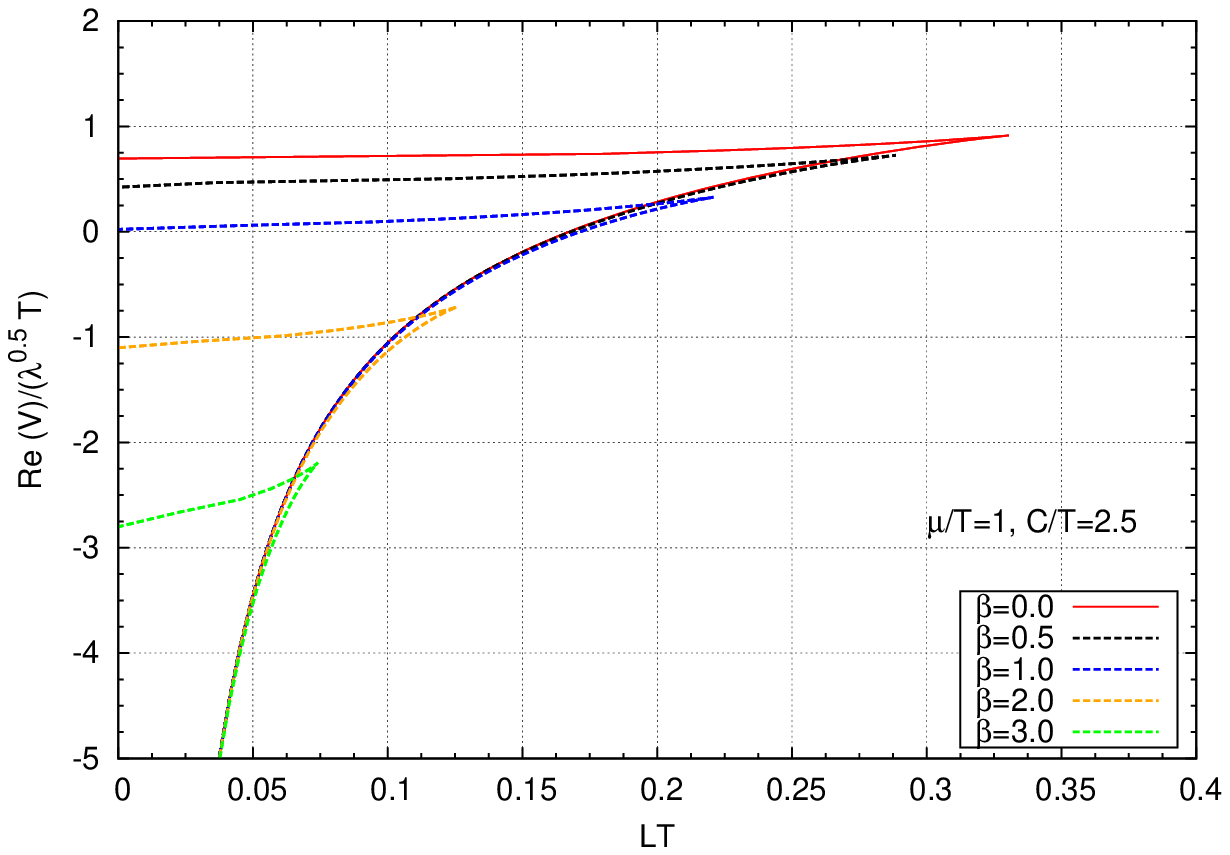}\\
		{\footnotesize perpendicular case}
	\end{minipage}
	\caption{Effect of the rapidity $\beta$ on the real $q\bar{q}$ potential.}\label{fig1}
	\end{figure}\\
In Fig.(\ref{fig1}), we probe the effect of rapidity $\beta$ on $\mathrm{Re}(V_{q\bar{q}})$. In both cases, we set the values $\frac{\mu}{T}=1$ and  $\frac{C}{T}=2.5$. For the parallel case, we see that the presence of the constraint on the turning-point $r_t$ of the U-shaped string profile restricts the domain for $\mathrm{Re}(V_{q\bar{q}})$ whenever $\beta\neq0$. In both cases, it is observed that with the increasing value of $\beta$, the value of $\mathrm{Re}(V_{q\bar{q}})$ decreases. It is also to be observed that for perpendicular scenario, the effect of rapidity on $\mathrm{Re}(V_{q\bar{q}})$ is more prominent.\\
\noindent The values of $LT_{max}$ in both cases are shows different, and it can be observed for any curve representing a particular value of $\beta$. For $\beta=1$ curve, the value of $LT_{max}$ $\approx 0.25$ for the parallel case and for the perpendicular case, $LT_{max} \approx 0.22$.\\
The curve for $\beta=0$ in both the plots represents the potential for heavy $q\bar{q}$ pair (at rest) in presence of the chemical potential $\mu$ and the warp factor $C$.\\
\begin{figure}[!h]
	\begin{minipage}[t]{0.48\textwidth}
		\centering\includegraphics[width=\textwidth]{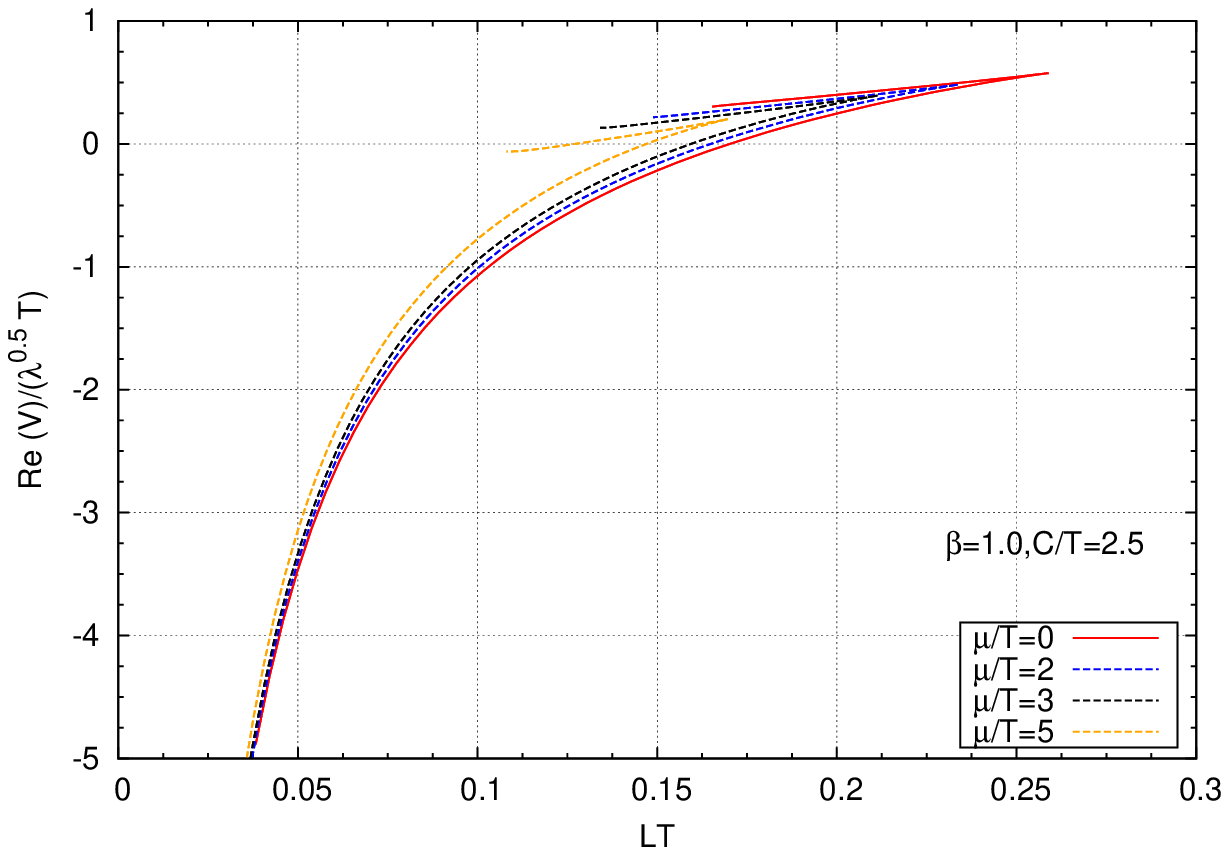}\\
		{\footnotesize parallel case}
	\end{minipage}\hfill
	\begin{minipage}[t]{0.48\textwidth}
		\centering\includegraphics[width=\textwidth]{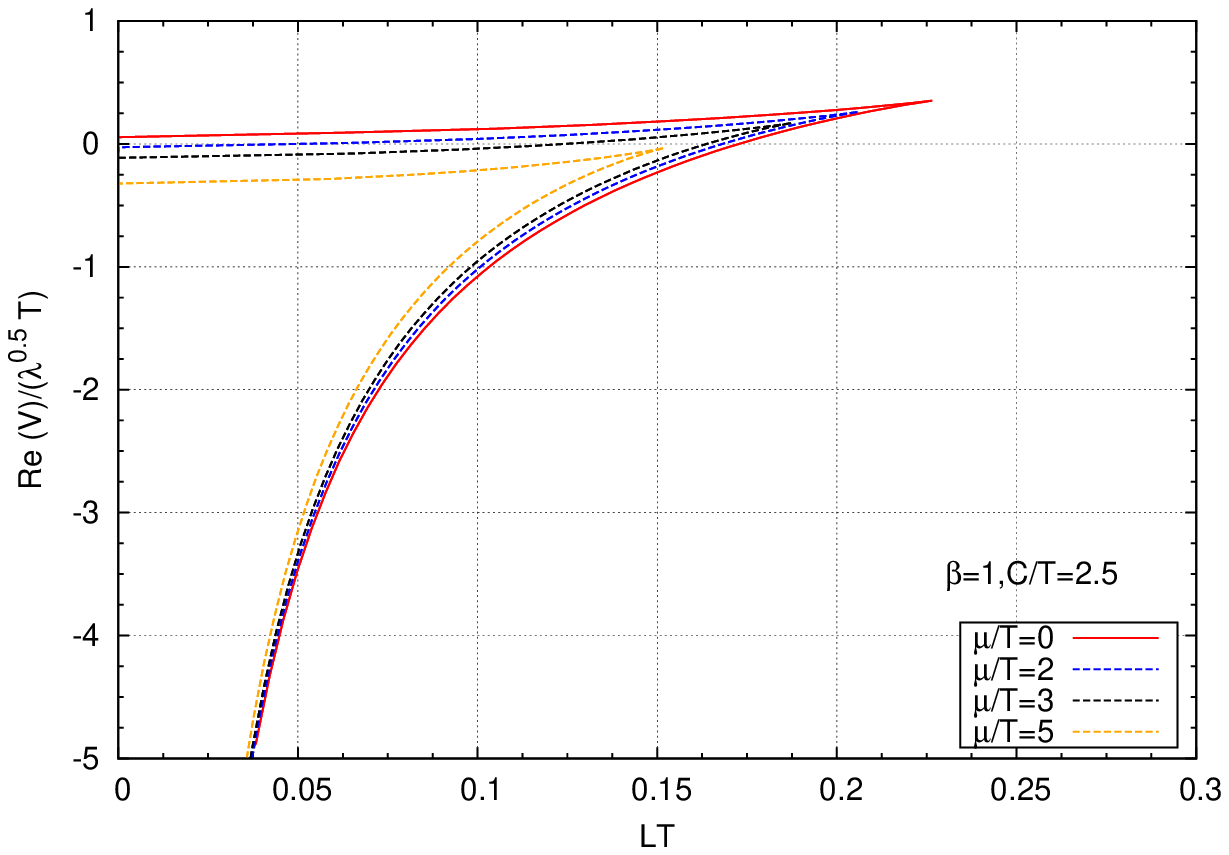}\\
		{\footnotesize perpendicular case}
	\end{minipage}
	\caption{Effect of the chemical potential $\mu$ on the real $q\bar{q}$ potential.}\label{fig2}
\end{figure}\\
\begin{figure}[!h]
	\begin{minipage}[t]{0.48\textwidth}
		\centering\includegraphics[width=\textwidth]{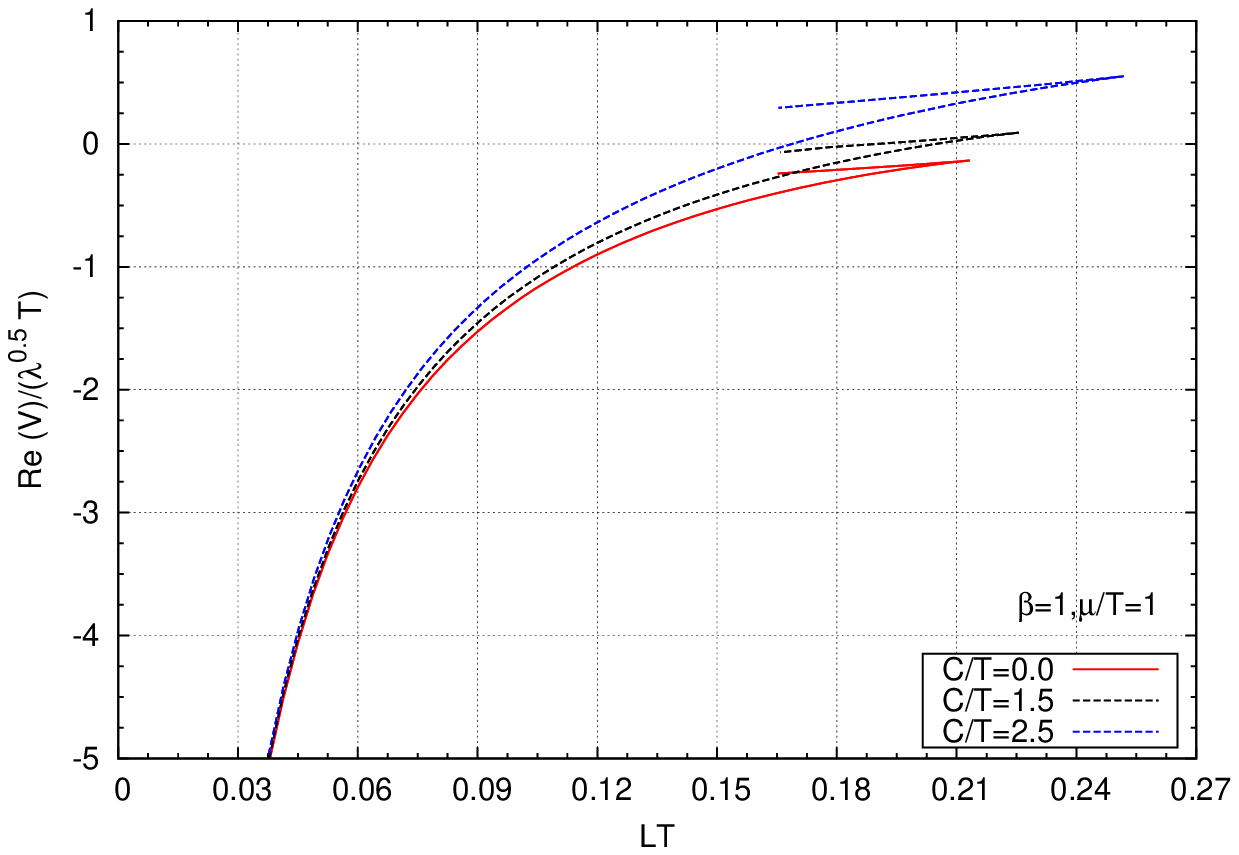}\\
		{\footnotesize parallel case}
	\end{minipage}\hfill
	\begin{minipage}[t]{0.48\textwidth}
		\centering\includegraphics[width=\textwidth]{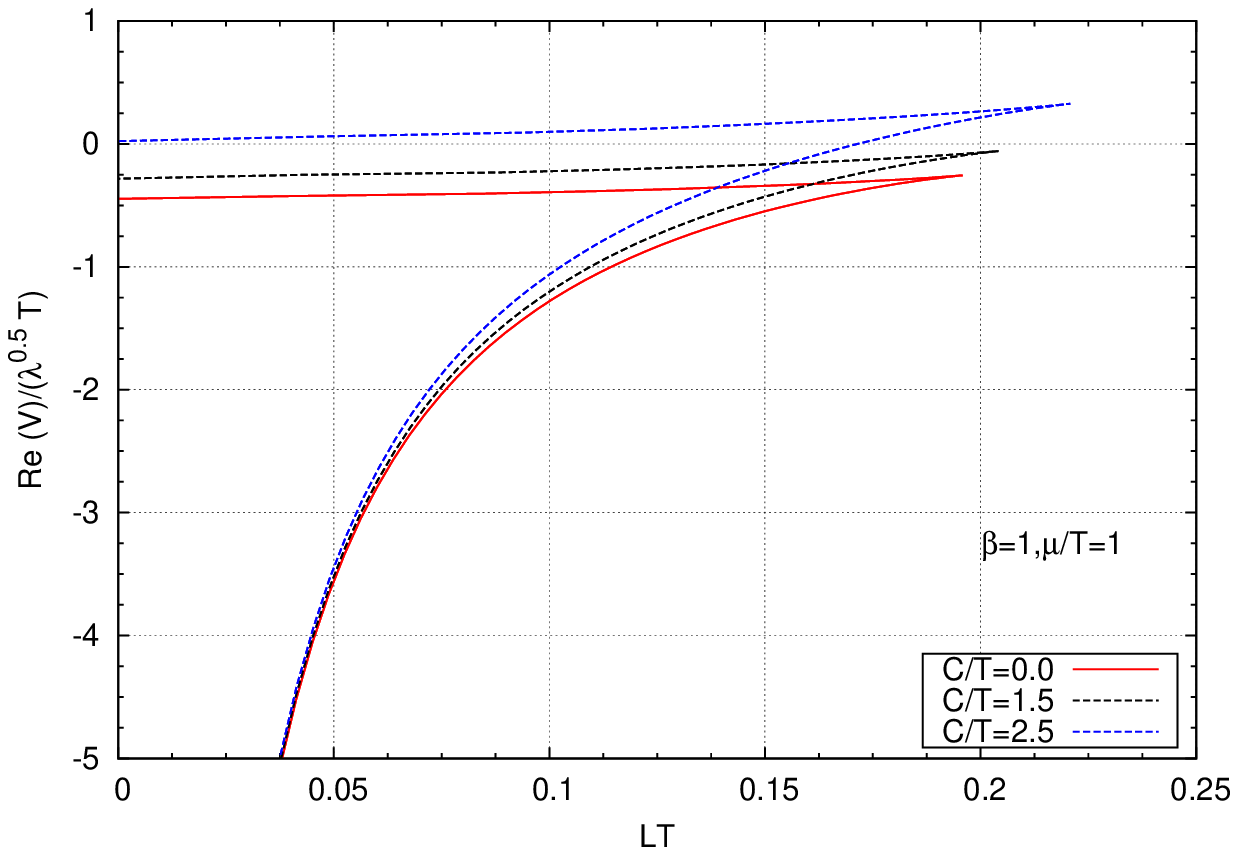}\\
		{\footnotesize  perpendicular case}
	\end{minipage}
	\caption{Effect of the nonconformality on the real $q\bar{q}$ potential.}\label{fig3}
\end{figure}
\noindent In Fig.(\ref{fig2}), we have shown the effect of the chemical potential $\mu$ on the real potential of a moving $q\bar{q}$ pair in the presence of nonconformality. It can be observed that, similar to the effect of rapidity $\beta$, with increase in $\mu$, the value of $\mathrm{Re}(V_{q\bar{q}})$ decreases. The constraint on the turning-point $r_t$ plays its role in the parallel case. The curve for $\frac{\mu}{T}=0$ represents the plasma without the chemical potential in both cases. It is also to be observed that in the perpendicular case, the value of $LT_{max}\approx 0.226$ for the curve $\frac{\mu}{T}=0$ but in case of the parallel scenario, $LT_{max}\approx 0.258$. By comparing Fig.(\ref{fig1}) and Fig.(\ref{fig2}), we can observe that, the value of $LT_{max}$ decreases more abruptly when we increase $\beta$ rather than the case when we increase $\mu$.
\noindent In Fig.(\ref{fig3}), we observe the effect of nonconformality on the real part of the $q\bar{q}$ potential. In both cases, the value of $LT_{max}$ increases with the increase in the value of the deformation parameter $C$. In $\mathrm{SW}_{T,\mu}$ with deformation $\frac{C}{T}=2.5$, the value of $LT_{max}$ $\approx 0.2518$ for the parallel case, and for the perpendicular case, $LT_{max} \approx 0.2207$.
\section{Jet quenching parameter ($q_m$) in the strong coupling limit of the field theory}\label{sec3}
We have seen that the presence of Lorentz boost has led to two different choices for the motion of the $q\bar{q}$ pair. Another fascinating phenomena can be observed in the limit $\beta\rightarrow\infty$, namely, the jet quenching of partons. Jet quenching represents the relativistic profile of the partons in strongly coupled plasma and is the property of the medium which probes the energy loss of the moving partons due to gluon radiation in the moving medium. In this limit ($\beta\rightarrow\infty$), the rectangular Wilson loop becomes light-like. This can also be done by writing down the metric in the lightcone coordinates \cite{liu1}. Note that the rapidity $\beta$ is related with the velocity as
\begin{eqnarray}
v = \tanh\beta~.
\end{eqnarray}
Hence in the limit $\beta\rightarrow\infty$, we have $v= 1$.
Before we proceed to calculate the jet quenching parameter, we perform a small simplication in the concerned spacetime metric. The metric coefficient $g_{tt}$ in the metric (\ref{boosted}) reads
\begin{eqnarray}
g_{tt} = -r^2 h(r)\bigg[f(r)\cosh^2{\beta} - \sinh^2{\beta}\bigg];~~ f(r)= 1-(1+Q^2) (\frac{r_h}{r})^4 +Q^2 (\frac{r_h}{r})^6\nonumber~.
\end{eqnarray}
In the computation of the jet quenching parameter ($q_m$), we need to take the limit $\beta\to\infty$ which can be confusing with the current expression for $g_{tt}$ in hand. This motivates us to simplify $g_{tt}$ as
\begin{eqnarray}\label{eq0}
g_{tt} = -r^2 h(r)\bigg[ \cosh^2\beta - \sinh^2\beta-\bigg[(1+Q^2) (\frac{r_h}{r})^4 -Q^2 (\frac{r_h}{r})^6\bigg]\cosh^2\beta\bigg] \equiv  -r^2 h(r)\bigg[1-a_Q(r)\cosh^2\beta\bigg]
\end{eqnarray}
where $a_Q(r)$ is given by
\begin{eqnarray} 
a_Q(r) = (1+Q^2) (\frac{r_h}{r})^4 -Q^2 (\frac{r_h}{r})^6~.
\end{eqnarray}
Now it can easily be observed from eq.(\ref{eq0}) that $g_{tt}\ll1$ in the limit $\beta \rightarrow \infty$. This leads to the open string action
\begin{eqnarray}
S_{NG} = \frac{i\tau}{\pi \alpha^{\prime}} \int_{-L/2}^{0} \mathcal{L}(r,r^{\prime})~ d\sigma
\end{eqnarray}
where the lagrangian $\mathcal{L}(r,r^{\prime})$ is a real quantity and reads 
\begin{eqnarray}
\mathcal{L}(r,r^{\prime})= r^2 h(r) \sqrt{\bigg(1+\frac{{r^{\prime}}^2}{r^4 f(r)}\bigg)\bigg(a_Q(r)\cosh^2\beta-1\bigg)}~.
\end{eqnarray}
Here we have used the fact that in the $\beta\rightarrow\infty$ limit, $a_Q(r)\cosh^2\beta>>1$.\\
\noindent The associated Hamiltonian reads
\begin{eqnarray}\label{jq1}
\mathcal{H}&=& \mathcal{L}(r,r^{\prime}) - r^{\prime} \frac{\partial\mathcal{L}(r,r^{\prime})}{\partial r^{\prime}}\nonumber\\
&=& r^2 h(r) \bigg[\frac{\sqrt{a_Q(r)\cosh^2\beta-1}}{\sqrt{1+\frac{{r^{\prime}}^2}{r^4 f(r)}}}\bigg]\nonumber\\
&=& \mathrm{constant} \equiv \gamma~.
\end{eqnarray}
Solving eq.(\ref{jq1}) for $r^{\prime}$ gives
\begin{eqnarray}
\frac{dr}{d\sigma}= \frac{r^4h(r) \sqrt{f(r)}}{\gamma} \sqrt{a_Q(r)\cosh^2\beta-1}\bigg[1- \frac{\gamma^2}{r^4 h^2(r) (a_Q(r)\cosh^2\beta-1)}\bigg]^{1/2}~.
\end{eqnarray}
To have real values of $r$ in the lage $\beta$ limit, the condition that must be satisfied reads
\begin{eqnarray}\label{4}
\gamma < r^2 h(r) \sqrt{a_Q(r)\cosh^2\beta-1}~.
\end{eqnarray}
Keeping the dominant contribution of $\gamma$, we obtain
\begin{eqnarray}\label{2}
\frac{dr}{d\sigma} \approx \frac{r^4 h(r) \sqrt{f(r)}}{\gamma} \sqrt{a_Q(r)\cosh^2\beta-1}~.
\end{eqnarray}
Integrating eq.(\ref{2}) we obtain,
\begin{eqnarray}\label{3}
L = 2 \gamma \int_{r_h}^{\infty} \frac{dr}{r^4 h(r) \sqrt{f(r)}\sqrt{a_Q(r)\cosh^2\beta-1}}~.
\end{eqnarray}
By using eq.(\ref{3}), we obtain the constant $\gamma$ in terms of the $q\bar{q}$ separation length $L$ to be
\begin{eqnarray}\label{6}
\gamma = \frac{L}{2} r_h^3  \bigg[\int_{0}^{1} \frac{u^2 ~du}{\exp((\frac{C}{T})^2 (\frac{u}{t_{\mu}})^2)\sqrt{f(u)}\sqrt{a_Q(u)\cosh^2\beta-1}}\bigg]^{-1}
\end{eqnarray}
where we have defined
\begin{eqnarray}
\frac{r_h}{r} &=& u \nonumber\\
f(u) &=& 1-\bigg[1+\frac{1}{3{t_{\mu}}^2}\bigg(\frac{\mu}{T}\bigg)^2\bigg] u^4 +\frac{1}{3{t_{\mu}}^2}\bigg(\frac{\mu}{T}\bigg)^2 u^6\nonumber\\
a_Q(u) &=& \bigg[1+\frac{1}{3{t_{\mu}}^2}\bigg(\frac{\mu}{T}\bigg)^2\bigg] u^4 -\frac{1}{3{t_{\mu}}^2}\bigg(\frac{\mu}{T}\bigg)^2 u^6 \nonumber~.
\end{eqnarray}
Substituting $\frac{dr}{d\sigma}$ in eq.(\ref{2}) into the Nambu-Goto action, we obtain
\begin{eqnarray}\label{5}
S_{NG} = \frac{i\tau}{\pi \alpha^{\prime}} \int_{r_h}^{\infty}dr \frac{h(r)\sqrt{a_Q(r)\cosh^2\beta-1}}{\sqrt{f(r)}}\times \bigg[1+\frac{\gamma^2}{r^4 h^2(r)(a_Q(r)\cosh^2\beta-1)}\bigg]^{1/2}~.
\end{eqnarray}
From the condition expressed in eq.(\ref{4}), we know that $\gamma$ is a small parameter and therefore we can expand the open string action given in eq.(\ref{5}) in powers of $\gamma$ as
\begin{eqnarray}
S_{NG} &\approx& \frac{i\sqrt{\lambda}\tau}{\pi} \int_{r_h}^{\infty} \frac{h(r)\sqrt{a_Q(r)\cosh^2\beta-1}}{f(r)} ~dr + \gamma^2 \frac{i\sqrt{\lambda}\tau}{2\pi}\int_{r_h}^{\infty} \frac{dr}{r^4 h(r)\sqrt{f(r)}\sqrt{a_Q(r)\cosh^2\beta-1}}+\mathcal{O}(\gamma^4)\nonumber\\
&\approx& S_0 + \gamma^2 S_I +...~.
\end{eqnarray}
It can be observed that in the limit $\gamma\to0$, the Nambu-Goto action $S_{NG}=S_0$ which is the self energy of two quarks moving in the plasma. $S_0$ can also be realized as the area of two disjoint world-sheets. Now subtracting the self energy term $S_0$ from the Nambu-Goto action, we obtain the regularized open string action to be
\begin{eqnarray}
S_{reg}&=&S_{NG} - S_0\approx\gamma^2 S_I\nonumber\\
 &=& \gamma^2 \frac{i\tau\sqrt{\lambda}}{2\pi r_h^3}\int_{0}^{1} \frac{u^2~du}{\exp((\frac{C}{T})^2 (\frac{u}{t_{\mu}})^2)\sqrt{f(u)}\sqrt{a_Q(u)\cosh^2\beta-1}}\nonumber\\
&=& \frac{iL^2\sqrt{\lambda}\tau}{8\pi}r_h^3 \bigg[\int_{0}^{1} \frac{u^2 ~du}{\exp((\frac{C}{T})^2 (\frac{u}{t_{\mu}})^2)\sqrt{f(u)}\sqrt{a_Q(u)\cosh^2\beta-1}}\bigg]^{-1}
\end{eqnarray}
where we have substituted the value of $\gamma$ from eq.(\ref{6}) in order to express the regularized action in terms of the $q\bar{q}$ separation length $L$. With the regularized action in hand, we now take the limit $\beta\to\infty$ to make the rectangular Wilson loop light-like. This yields
\begin{eqnarray}
S_{reg} &=& \frac{iL^2\sqrt{\lambda}r_h^3}{8\pi}~(\tau\cosh\beta) \left[\int_{0}^{1} \frac{u^2}{\exp((\frac{C}{T})^2 (\frac{u}{t_{\mu}})^2)\sqrt{f(u)}\sqrt{a_Q(u)}}\times\frac{1}{\sqrt{1-\frac{1}{a_Q(u)\cosh^2\beta}}}~du\right]^{-1}\nonumber\\
&\approx& \frac{iL^2\sqrt{\lambda}r_h^3}{8\pi}~(\tau\cosh\beta) \left[\int_{0}^{1} \frac{u^2}{\exp((\frac{C}{T})^2 (\frac{u}{t_{\mu}})^2)\sqrt{f(u)}\sqrt{a_Q(u)}}~du\right]^{-1}
\end{eqnarray}
where we keep only the dominating term in $\beta$.\\
By identifying the transverse screening length of a moving $q\bar{q}$ pair as \cite{liu2}
\begin{eqnarray}
	\tau\cosh\beta = \frac{L^{-}}{\sqrt{2}}
\end{eqnarray}
we write down the regularized action in the final form as
\begin{eqnarray}\label{7}
S_{reg} = \frac{iL^{-}L^2\sqrt{\lambda}r_h^3}{8\sqrt{2}\pi}~ \bigg[\int_{0}^{1} \frac{u^2 ~du}{\exp((\frac{C}{T})^2 (\frac{u}{t_{\mu}})^2)\sqrt{f(u)}\sqrt{a_Q(u)}}\bigg]^{-1}~.
\end{eqnarray}
The standard relation between the regularized string world-sheet area and the jet quenching parameter $q_m$ reads \cite{liu1}
 \begin{eqnarray}\label{8}
 \langle W[\mathcal{C}]\rangle = e^{2iS_{reg}} \approx e^{-\frac{q_m}{4\sqrt{2}}L^{-}L^2}~.
 \end{eqnarray}
By substituting eq.(\ref{7}) in eq.(\ref{8}), we obtain the jet quenching parameter $q_m$ to be
\begin{eqnarray}\label{11}
q_m &=& \frac{\sqrt{\lambda}r_h^3}{\pi}~ \bigg[\int_{0}^{1} \frac{u^2 ~du}{\exp((\frac{C}{T})^2 (\frac{u}{t_{\mu}})^2)\sqrt{f(u)}\sqrt{a_Q(u)}}\bigg]^{-1}\nonumber\\
&=&  \frac{\sqrt{\lambda}}{\pi}\bigg[\frac{\pi}{2}+ \sqrt{\frac{\pi^2}{4}+\frac{1}{6}\big(\frac{\mu}{T}\big)^2}\bigg]^3~T^3 \bigg[\int_{0}^{1} \frac{u^2 ~du}{\exp((\frac{C}{T})^2 (\frac{u}{t_{\mu}})^2)\sqrt{f(u)}\sqrt{a_Q(u)}}\bigg]^{-1}~.
\end{eqnarray}
The above expression for $q_m$ is given completely in terms of the boundary field theoretical parameters which are temperature $T$, chemical potential $q$ and warp factor $C$. In the limit $\frac{\mu}{T}\to 0$ and $\frac{C}{T}\to 0$, we obtain
\begin{eqnarray}\label{23}
q_m = \pi^{3/2}\frac{\Gamma[\frac{3}{4}]}{\Gamma[\frac{5}{4}]}\sqrt{\lambda}T^3
\end{eqnarray} 
which is the well-known jet quenching parameter $q_m$ of a $q\bar{q}$ pair in a $\mathcal{N}=4$ super Yang-Mills plasma \cite{liu1}.
\begin{figure}[!h]
	\begin{minipage}[t]{0.48\textwidth}
		\centering\includegraphics[width=\textwidth]{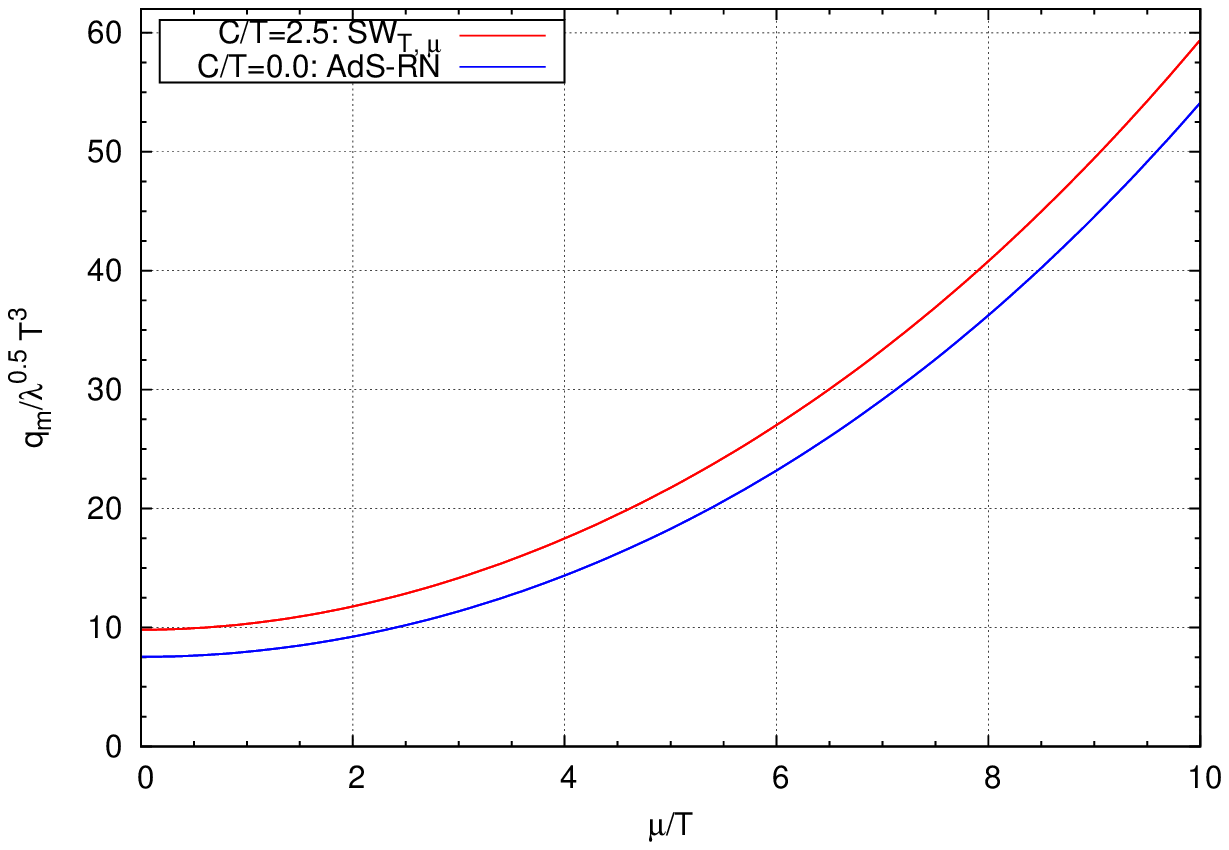}\\
		{\footnotesize Variation of jet quenching parameter with respect to $\frac{\mu}{T}$}
	\end{minipage}\hfill
	\begin{minipage}[t]{0.48\textwidth}
		\centering\includegraphics[width=\textwidth]{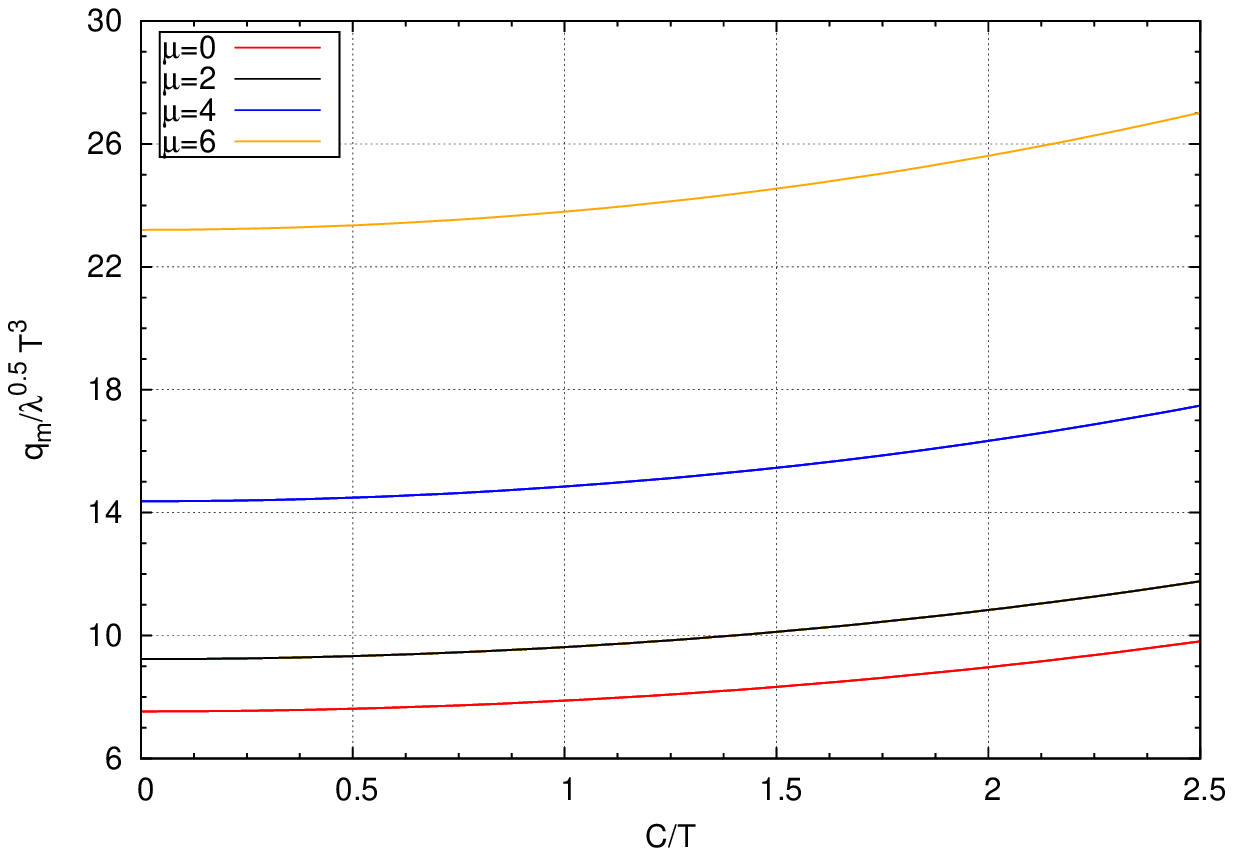}\\
		{\footnotesize Influence of nonconformality on jet quenching parameter}
	\end{minipage}
	\caption{Behaviour of the $q_m$ parameter in presence of nonconformality and chemical potential.}
	\label{fig6}
\end{figure}\\
\noindent The plot in the left panel of Fig.(\ref{fig6}) represents the behaviour of the jet quenching parameter $q_m$ with respect to the chemical potential $\mu$ for both with and without nonconformality. It is to be observed that in both cases the value of $q_m$ increases with the increase in the value of the chemical potential. In the right panel of Fig.(\ref{fig6}), we depict the effect of nonconformality on the jet quenching for a chosen value of the chemical potential. It is observed that for lower value of the deformation parameter $\frac{C}{T}$ the value of jet quenching remains almost same. However, at higher values of the deformation parameter ($\frac{C}{T}>1$), the effect on the jet quenching is much more pronounced and similar to the effect of the chemical potential, it also increases the amount of in medium energy loss. The above plots suggest that the presence of non-zero chemical potential and nonconformality enhances the in-medium energy loss in the high $p_T$ region which can be tested via the RHIC experiments. In the limit $\frac{\mu}{T}\rightarrow0$, the above discussion qualitatively matches with the results obtained in \cite{robust}. Recent observations made in RHIC shows that after collision, QGP expands and the jet quenching decreases with decrease in temperature and the observed values for the jet quenching parameter is $5-15~$GeV$^2$/fm \cite{rhic1}-\cite{rhic3}. We can compare our results with this observed value for $q_m$. In the left panel of Fig.(\ref{fig8}), we plot the jet quenching parameter with respect to the temperature at the conformal limit ($C\rightarrow 0$). From this we observe that our result agrees with RHIC in the range $0.62\leq T\leq 0.9$ GeV for $\mu=0$ GeV, $0.59\leq T\leq 0.88$ GeV for $\mu= 1$ GeV and $0.5\leq T\leq 0.81$ GeV for $\mu= 2$ GeV. This observation in turn reveals that increase in the finite quark density in the strongly coupled plasma produces the observed values of jet quenching at relatively lower value of the temperature region. In the right panel of Fig.(\ref{fig8}), we plot it for $C=2.5\times~$T GeV. Similar to the effect of the chemical potential $\mu$, deformation parameter $C$ (nonconformality) further lowers the required temperature region in order to observe the obtained RHIC values for jet quenching.  
\begin{figure}[!h]
	\begin{minipage}[t]{0.48\textwidth}
		\centering\includegraphics[width=\textwidth]{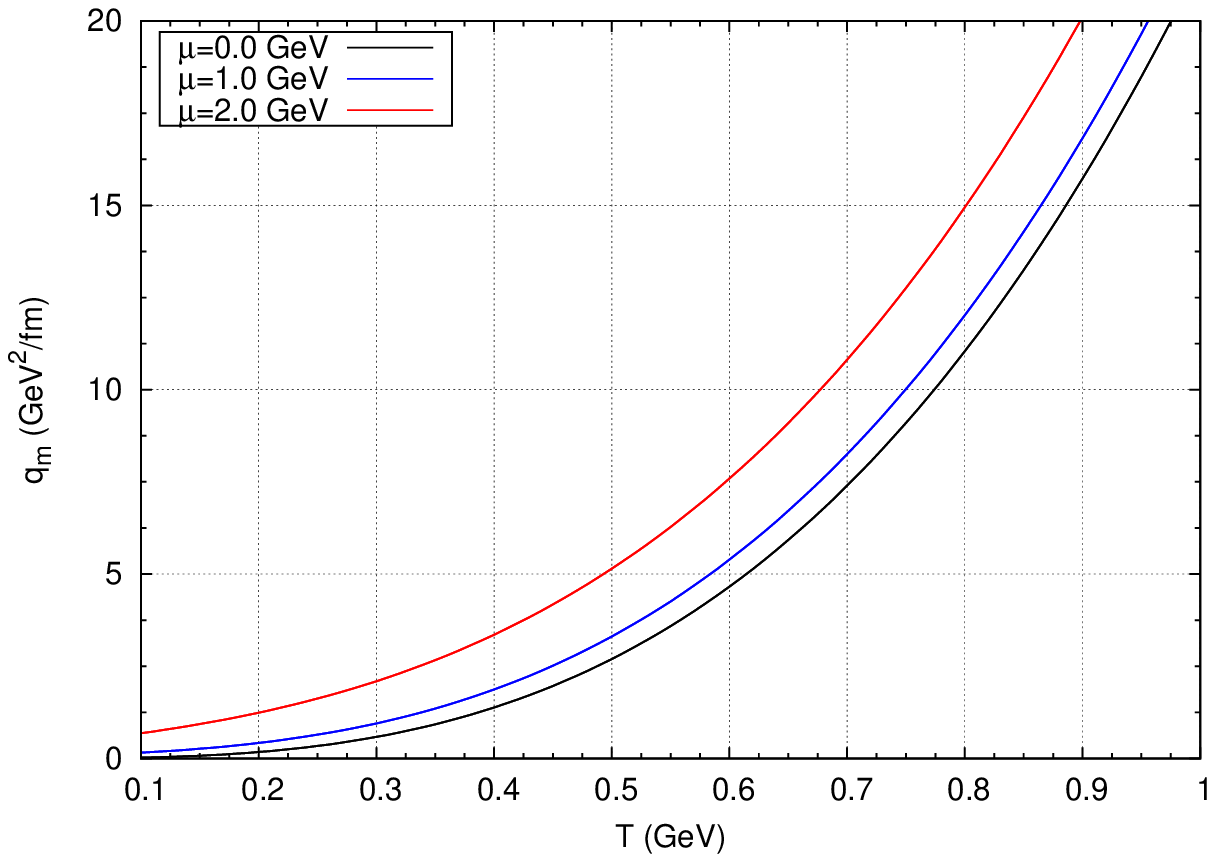}\\
		{\footnotesize $C=0$}
	\end{minipage}\hfill
	\begin{minipage}[t]{0.48\textwidth}
		\centering\includegraphics[width=\textwidth]{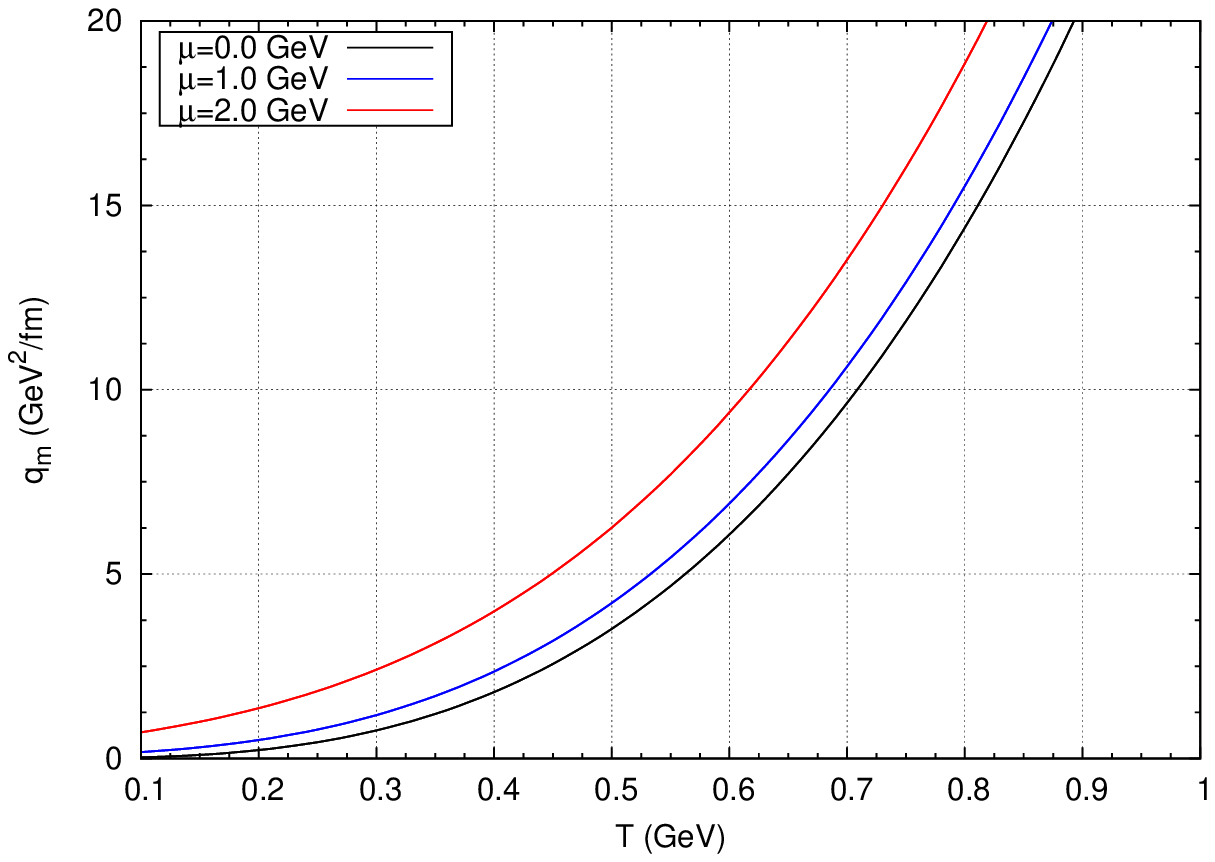}\\
		{\footnotesize $C=2.5\times~$T GeV}
	\end{minipage}
	\caption{Jet-quenching parameter $q_m$ vs temperature at various values of $\mu$.}
	\label{fig8}
\end{figure}\\
We now compare our result of $q_m$ (given in eq.(\ref{11})) with the observed values of the jet quenching parameter in RHIC and LHC at fixed temperatures. The observed values of $q_m$ in the most central Au-Au collisions at RHIC reads $q_m = 1.2 \pm 0.3$ GeV$^2$/fm at the highest temperature $T = 0.37$ GeV and in the most central Pb-Pb collisions at LHC reads $q_m=1.9\pm0.7$ GeV$^2$/fm at the highest temperature $T=0.47$ GeV \cite{LHC}. We now use the above values of temperature $T$ in eq.(\ref{11}) and constraint the values of the chemical potential $\mu$ and the deformation parameter $C$ in such a way that the resulting values of $q_m$ satisfies the experimentally observed values mentioned above.
\begin{figure}[!h]
	\begin{minipage}[t]{0.48\textwidth}
		\centering\includegraphics[width=\textwidth]{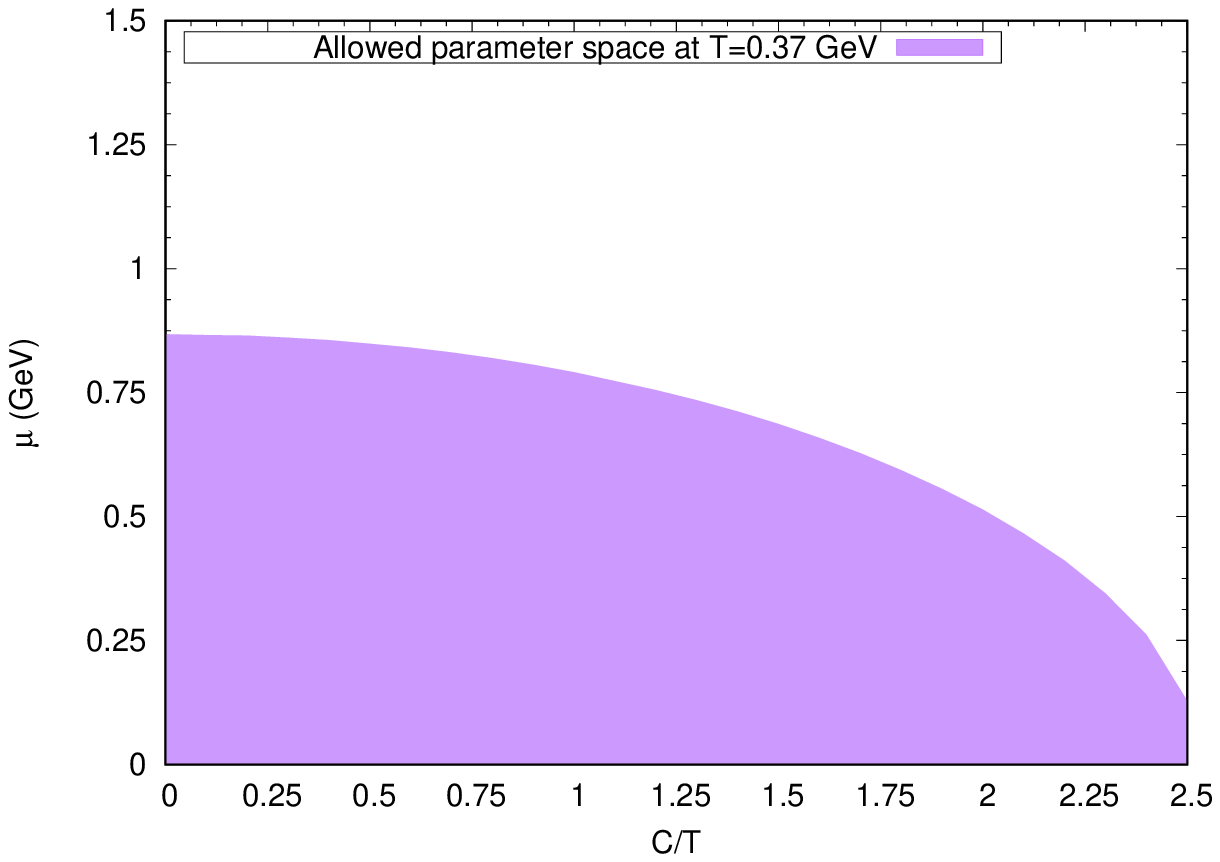}\\
		{\footnotesize $T=0.37$ GeV and $q_m=1.2\pm0.3$ GeV$^2$/fm}
	\end{minipage}\hfill
	\begin{minipage}[t]{0.48\textwidth}
		\centering\includegraphics[width=\textwidth]{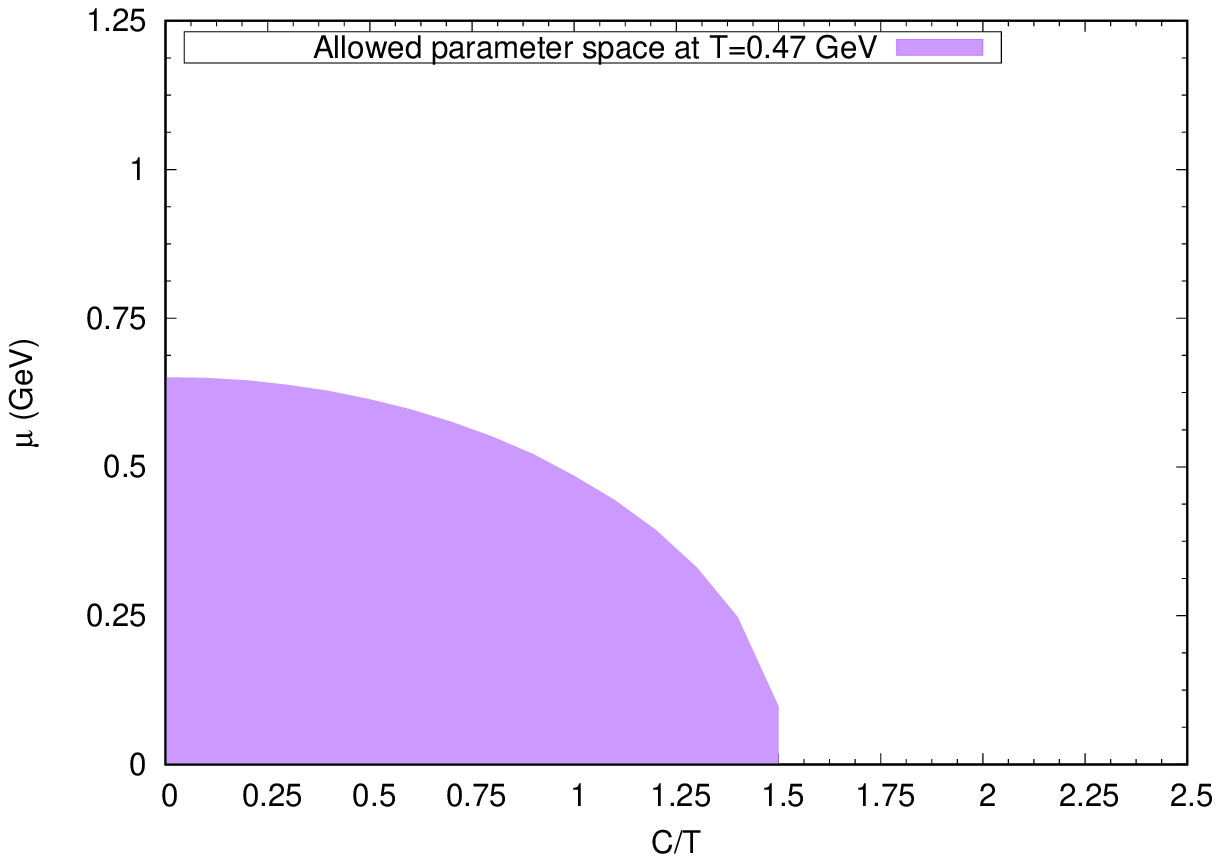}\\
		{\footnotesize $T=0.47$ GeV and $q_m=1.9\pm0.7$ GeV$^2$/fm}
	\end{minipage}
	\caption{Allowed values of $\mu$ and $C$ corresponding to the experimentally observed values of $q_m$ at RHIC and LHC.}
	\label{paramspace}
\end{figure}\\
In Fig.(\ref{paramspace}), we graphically represent the allowed values of $\mu$ and $C$ at the mentioned temperatures. The left plot corresponds to the value of the jet quenching parameter $q_m=1.2\pm0.3$ GeV$^2$/fm observed at the RHIC at temperature $T=0.37$ GeV and the right plot corresponds to the value of the jet quenching parameter $q_m=1.9\pm0.7$ GeV$^2$/fm observed at the LHC at temperature $T=0.47$ GeV. Note that the allowed values in the parameter space of $\mu$ and $C$ shrinks when the temperature increases. We can also compute theoretical values of $q_m$ and compare it with the experimentally observed values. In particular, for $\mu=0$ with $0\leq \frac{C}{T} \leq 2.5$ at $T=0.37$ GeV, the value of the jet quenching parameter obtained from eq.(\ref{11}) is found to be $q_m=1.31\pm 0.18$ GeV$^2$/fm.\\
\noindent Another thing that we would like to discuss in this section is to exploit the relation between $q_m$ and shear viscosity to entropy density ratio ($\frac{\eta}{s}$) in order to probe the strength of the coupling in the strong coupling limit. In \cite{eq}, it is conjectured that in the strong coupling limit (large $N$ limit), the shear viscosity to entropy density ratio shows the following behaviour
\begin{eqnarray}\label{10}
\frac{\eta}{s} \gg \mathcal{K}~.
\end{eqnarray}
On the other hand in the weakly coupled limit (quasiparticle dominated quark-gluon plasma)
\begin{eqnarray}
 \frac{\eta}{s} \approx  \mathcal{K};~\mathcal{K}=1.25 \frac{T^3}{q_m}~.
\end{eqnarray}
The above relations suggest that the quantity $\mathcal{K}$ can be treated as a order parameter which probes the behaviour of the coupling constant of the gauge theory as the gauge theory must obey the condition $\frac{\eta}{s} \gg  \mathcal{K}$ in order to be a strongly coupled one. Interestingly, small values of $\eta/s$ is also favoured by hydrodynamic models describing elliptic flows observed in experiments involving RHIC \cite{ebs}. This observation was supported from AdS/CFT calculations where it was shown that the ratio of the shear viscosity to entropy density is given by $\eta/s=\frac{\hbar}{4\pi k_B}$, which turns out to be indeed a small value. This result obtained in the AdS/CFT set up led to the speculation that for all relativistic quantum field theories at finite temperature and zero chemical potential has a lower bound \cite{ebsbound,ebsbound2}
 \begin{eqnarray}\label{ebsbound}
 \eta/s \geq \frac{\hbar}{4\pi k_B}~.
 \end{eqnarray}
 As mentioned above, AdS/CFT computations saturated the above bound. The above conjecture is also supported in the strongly coupled limit of $\mathcal{N}=4$ supersymmetric SU(N) Yang-Mills theory, where it has been shown that $\eta/s$ has the well known form \cite{N=4sym} 
\begin{eqnarray}\label{24}
\frac{\eta}{s} = \frac{1}{4\pi}\bigg[1+\frac{135 \xi(3)}{\lambda^{3/2}}+...\bigg]~.
\end{eqnarray}
 \\
By substituting the jet quenching parameter ($q_m$) of strongly coupled $\mathcal{N}=4$ SYM (given in eq.(\ref{23})) in eq.(\ref{10}), one obtains \cite{eq}
\begin{eqnarray}
\frac{0.166}{\sqrt{\lambda}}\ll\frac{\eta}{s}~.
\end{eqnarray}
From the above relation we find that for $\lambda\gg 4.3537$, the strongly coupled behaviour of the finite temperature quantum field theory is sustained.
On a similar note, we substitute our result for $q_m$ (given in eq.(\ref{11})) in eq.(\ref{10}) and obtain
\begin{eqnarray}
\frac{\eta}{s} \gg 1.25\frac{\pi}{\sqrt{\lambda}} \frac{1}{\bigg[\frac{\pi}{2}+ \sqrt{\frac{\pi^2}{4}+\frac{1}{6}(\frac{\mu}{T})^2}\bigg]^3}\int_{0}^{1} \frac{u^2 ~du}{\exp((\frac{C}{T})^2 (\frac{u}{t_{\mu}})^2)\sqrt{f(u)}\sqrt{a_Q(u)}}~.
\end{eqnarray}
Once again making use of the conjecture about the lower bound of $\eta/s$ (given in eq.(\ref{ebsbound})) for all relativistic finite temperature quantum field theories, we get $\lambda\gg 2.5632$ for $\frac{\mu}{T}= 0$ with the deformation parameter $\frac{C}{T}=2.5$. This suggests that if the conformal invariance in the gauge theory is broken, the minimum value for the t'Hooft coupling constant $\lambda$ reduces with increasing $\frac{C}{T}$.

\section{Imaginary part of the $q\bar{q}$ pair potential}\label{sec4}
In this section, we compute the imaginary part of the $q\bar{q}$ pair which arises due to the thermal fluctuation $\delta r(x)$ of the world-sheet around the classical string configuration $r_c(x)$ at some finite temperature \cite{im1}. The fluctuations are of the form
\begin{eqnarray}
r(x) = r_c(x) + \delta r(x)
\end{eqnarray}
with the boundary conditions $\delta r(x=\pm\frac{L}{2})=0$. For the sake of simplicity, we assume the fluctuations $\delta r(x)$ is to be of arbitrary long wavelength, that is $\frac{d\delta r(x)}{dx}\rightarrow 0$.
In the semiclassical approximation, the string partition function taking into account the fluctuation reads
\begin{eqnarray}\label{19}
Z_{string} \approx \int \mathcal{D}\delta r(x) e^{i S_{NG}(r_c(x)+\delta r(x))}~. 
\end{eqnarray}
We now discretize the above string partition function in the limit $-\frac{L}{2}\leq x\leq \frac{L}{2}$ by considering $2N$ points $x_k = k \Delta x$ with the definition $k = -N,~-N+1,...,+N$ and $\Delta x= \frac{L}{2N}$. By using these, we obtain
\begin{eqnarray}\label{12}
Z_{string} \approx \lim\limits_{N \to \infty} \int d[\delta(x_{-N})]...d[\delta(x_{+N})] \exp(\frac{i\tau \Delta x}{2\pi\alpha^{\prime}}\sum_{k} \sqrt{A(r_k) + B(r_k) {r_k^{\prime}}^2}~)
\end{eqnarray}
where $r_k\equiv r(x_k)$ and $r_k^{\prime} \equiv r^{\prime}(x_k)$. The thermal fluctuations are more prominent around $x=0$ where $r=r_t$. The turning point $r=r_t$ denotes the closest point to the event horizon of the U-shaped string profile. Keeping in mind the importance of the turning point $r=r_t$, we expand $r_c(x_k)$ around $x=0$. Keeping terms upto second order in $x_k$, we get
\begin{eqnarray}
	r_c(x_k) &\approx& r_c(x=0) + r_c^{\prime}(x=0)x_k + \frac{1}{2} r_c^{\prime\prime}(x=0)x_k^2 \nonumber\\
	&\approx& r_t + \frac{1}{2} r_c^{\prime\prime}(x=0)x_k^2
\end{eqnarray} 
where we have used the following definitions of the turning point $r_t$
\begin{eqnarray}
	r_c(x=0)\equiv r_t;~~ r_c^{\prime}|_{r_c=r_t} =0~.
\end{eqnarray}
On the other hand, the expansions for $A(r_k)$ and $B(r_k)$ yields
\begin{eqnarray}
A(r_k) &\approx& A_t + A_t^{\prime} \delta r+ \frac{1}{2} A_t^{\prime}r_c^{\prime \prime}(x=0)x_k^2+\frac{1}{2}A_t^{\prime\prime}\delta r^2\nonumber\\
B(r_k) &\approx& B_t\nonumber
\end{eqnarray}
where $A_t\equiv A(r_t)$ and $B_t \equiv B(r_t)$. In the above expansions we have kept terms upto second order in $x_k^m\delta r_n$ ($m+n\leq2$). Substituting the above expanded forms in the Nambu-Goto action, given in the exponential of eq.($\ref{12}$), we obtain
\begin{eqnarray}
S_{NG}^k &\equiv& \frac{\tau \Delta x}{2\pi\alpha^{\prime}}\sqrt{A(r_k) + B(r_k) {r_k^{\prime}}^2} \nonumber\\
 &=& \frac{\tau\Delta x}{2\pi\alpha^{\prime}}\sqrt{D_1 x_k^2+D_2}
\end{eqnarray}
where
\begin{eqnarray}\label{13}
D_1 &=& \frac{r_c^{\prime\prime}(0)}{2}[2B_t r_c^{\prime\prime}(x=0)+A_t^{\prime}]\nonumber\\
D_2 &=& A_t +\delta r A_t^{\prime} + \frac{1}{2}\delta r^2 A_t^{\prime\prime}\nonumber.
\end{eqnarray}
It is to be noted that if the function inside the square root of the above string action is negative then it will contribute to the imaginary part of the $q\bar{q}$ potential. The relevant domain for the integral in the parition function will be defined by the roots of the function inside the square root which is given in eq.(\ref{13}). This leads to the $k^{th}$ component of the integral given as
\begin{eqnarray}\label{14}
	I_{k} \equiv \int_{\delta r_k^{min}}^{\delta r_k^{max}} d(\delta r_k) \exp\left(\frac{i\tau \Delta x}{2\pi\alpha^{\prime}}\sqrt{D_1 x_k^2+D_2}\right)
\end{eqnarray}
where $\delta r_k^{min}$ and $ \delta r_k^{max}$ are the roots of $D_1 x_k^2+D_2$ in $\delta r$.
The above integral can be solved by saddle point method for $\alpha^{\prime}\ll1$. The function in the exponential has a stationary point when the function $
D_1 x_k^2+D_2$ assumes an extremal value which happens for $\delta r = - \frac{A_t^{\prime}}{A_t^{\prime\prime}}$. This further simplifies 
\begin{eqnarray}
	D_1 x_k^2+D_2 \equiv D_1 x_k^2 + A_t -\frac{{A_t^{\prime}}^2}{2A_t^{\prime\prime}}~.
\end{eqnarray}
Now at $\delta r_k^{max}$ and $\delta r_k^{min}$ we have
\begin{eqnarray}
	 D_1 x_k^2 + A_t -\frac{{A_t^{\prime}}^2}{2A_t^{\prime\prime}} = 0
\end{eqnarray}
which leads to 
\begin{eqnarray}
x_k = \sqrt{\frac{1}{D_1}\left(\frac{{A_t^{\prime}}^2}{2A_t^{\prime\prime}}-A_t\right)} \equiv x_c.
\end{eqnarray}
The total contribution to the imaginary part comes from $\Pi_k I_k$, yielding
\begin{eqnarray}\label{15}
\mathrm{Im}(V_{q\bar{q}}) = - \frac{1}{2\pi\alpha^{\prime}} \int_{|x|<x_c} dx \sqrt{-x^2D_1-A_t+\frac{{A_t^{\prime}}^2}{2A_t^{\prime\prime}}}~.
\end{eqnarray}
Integrating eq.(\ref{15}), we obtain
\begin{eqnarray}\label{18}
\mathrm{Im}(V_{q\bar{q}}) = - \frac{1}{2\sqrt{2}\alpha^{\prime}} \sqrt{B_t}\bigg(\frac{A_t^{\prime}}{2A_t^{\prime\prime}}-\frac{A_t}{A_t^{\prime}}\bigg)~.
\end{eqnarray}

\noindent We now introduce $\frac{r_h}{r_t} = \alpha,~\frac{1}{\alpha^{\prime}}=\sqrt{\lambda}$ and write down the imaginary part of the $q\bar{q}$ potential in a dimensionless form as
\begin{eqnarray}\label{17}
\frac{\mathrm{Im}(V_{q\bar{q}})}{\sqrt{\lambda}T} = - \frac{1}{2\sqrt{2}} \sqrt{B(\alpha)}\bigg(\frac{A^{\prime}(\alpha)}{2A^{\prime\prime}(\alpha)}-\frac{A(\alpha)}{A^{\prime}(\alpha)}\bigg)
\end{eqnarray}
The expressions for $A(\alpha),~A^{\prime}(\alpha),~ A^{\prime\prime}(\alpha)$ and $B(\alpha)$ are given in the appendix for both the parallel and the perpendicular cases.
\subsection{Numerical analysis and observations}
We shall now proceed to numerical analysis in order to observe the effects of rapidity $\beta$, deformation paramter $\frac{C}{T}$ ans chemical potential $\frac{\mu}{T}$ on the imaginary part of the $q\bar{q}$ potential. 
 \begin{figure}[!h]
 	\begin{minipage}[t]{0.48\textwidth}
 		\centering\includegraphics[width=\textwidth]{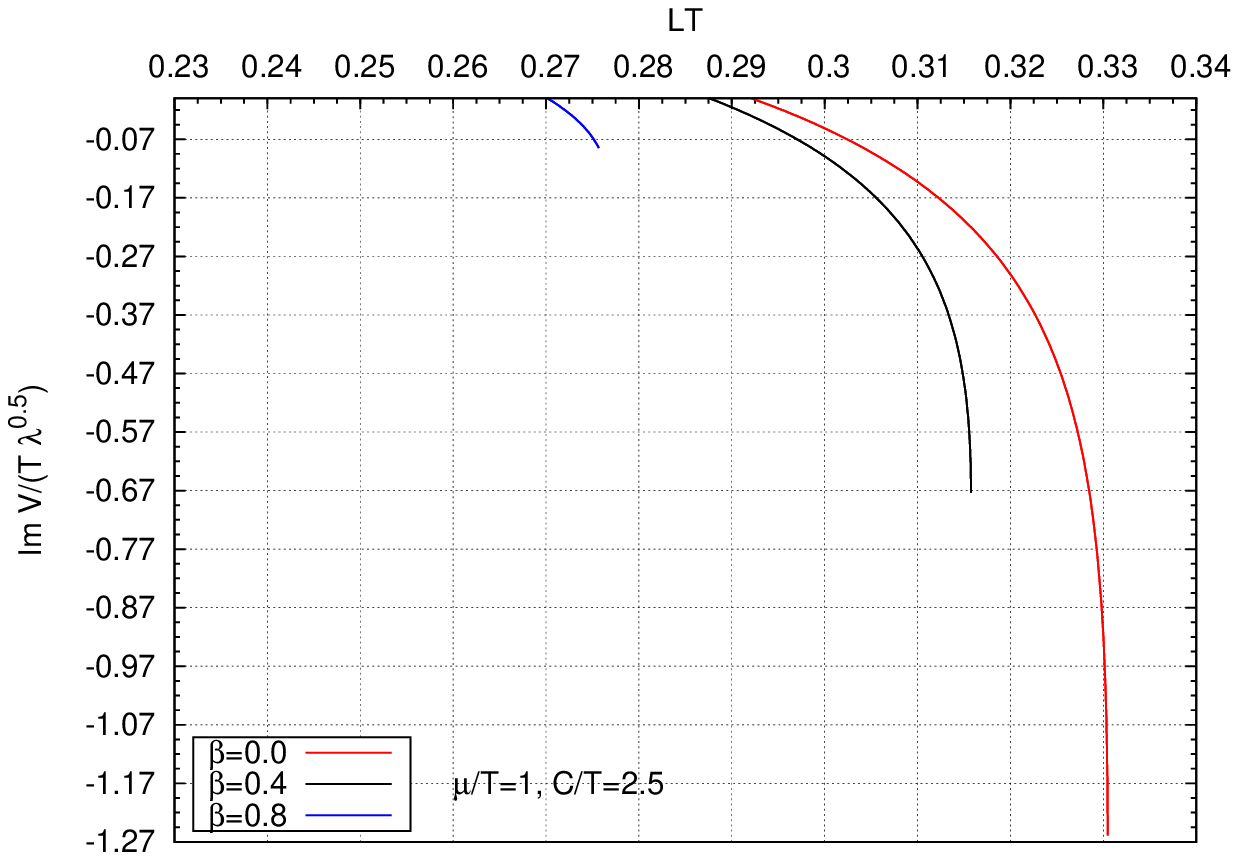}\\
 		{\footnotesize parallel case}
 	\end{minipage}\hfill
 	\begin{minipage}[t]{0.48\textwidth}
 		\centering\includegraphics[width=\textwidth]{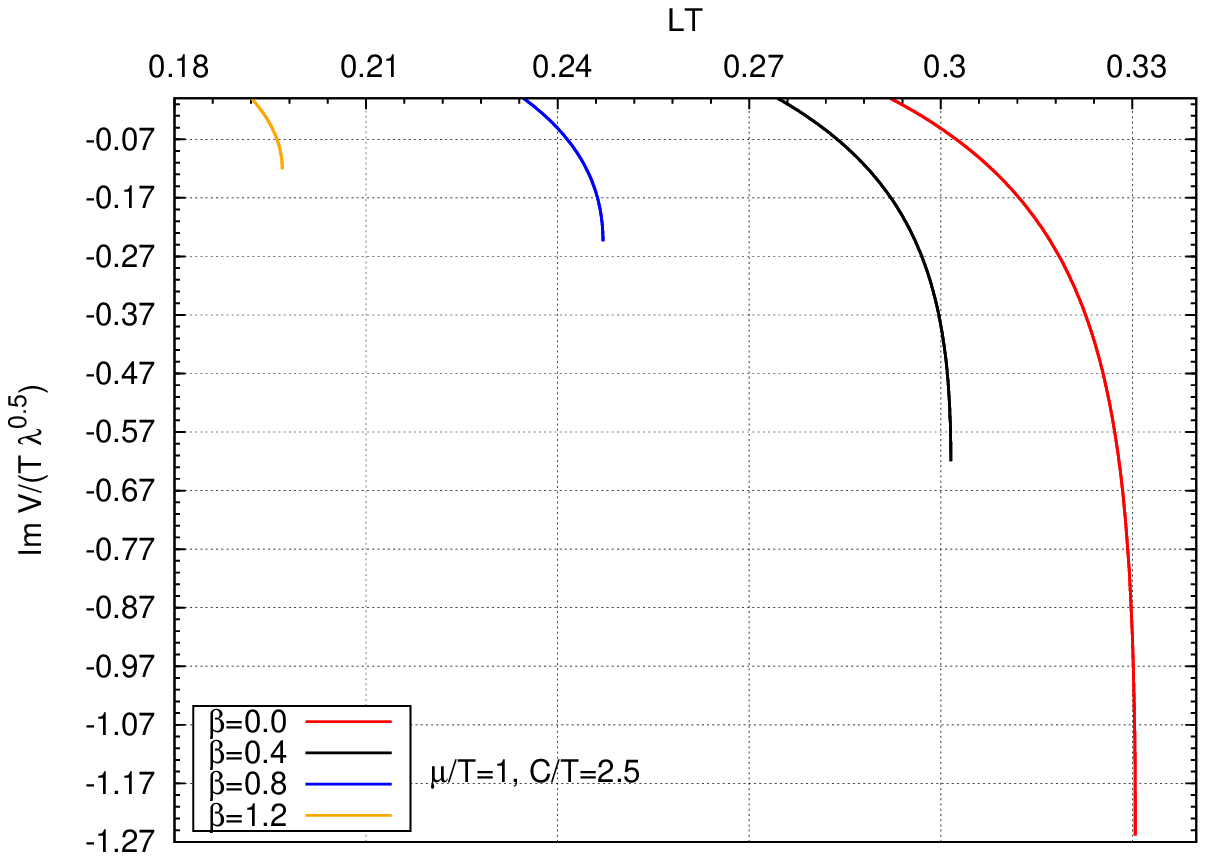}\\
 		{\footnotesize perpendicular case}
 	\end{minipage}
 	\caption{Effect of rapidity $\beta$ on the imaginary $q\bar{q}$ potential (we set $\frac{C}{T} =2.5, ~\frac{\mu}{T}=1$).}\label{fig5}
 \end{figure}\\
Fig.(\ref{fig5}) depicts the effect of the Lorentz boost on the imaginary potential for different orientation of the $q\bar{q}$ pair. In both panels, the curve for $\beta=0$ represents the static $q\bar{q}$ pair which is same in both of the plots and also in both cases imaginary potential starts from a smaller value of $LT$ when the rapidity $\beta$ increases. However, it is observed that $ \mathrm{Im}(V_{q\bar{q}})=0$ for  $\beta = 1.2$ in the parallel case, whereas $\mathrm{Im}(V_{q\bar{q}})\neq0$ in the perpendicular case for the same value of $\beta$. This suggests that in case of the parallel orientation of the $q\bar{q}$ pair, thermal width vanishes at smaller value of $\beta$ compared to the transverse orientation of the $q\bar{q}$ pair. By comparing both of the above plots, it is observed that the anisotropy introduced in the analysis via Lorentz boost creates a strong suppression of the thermal width at smaller angles in the strong coupling limit.

\begin{figure}[!h]
	\begin{minipage}[t]{0.48\textwidth}
		\centering\includegraphics[width=\textwidth]{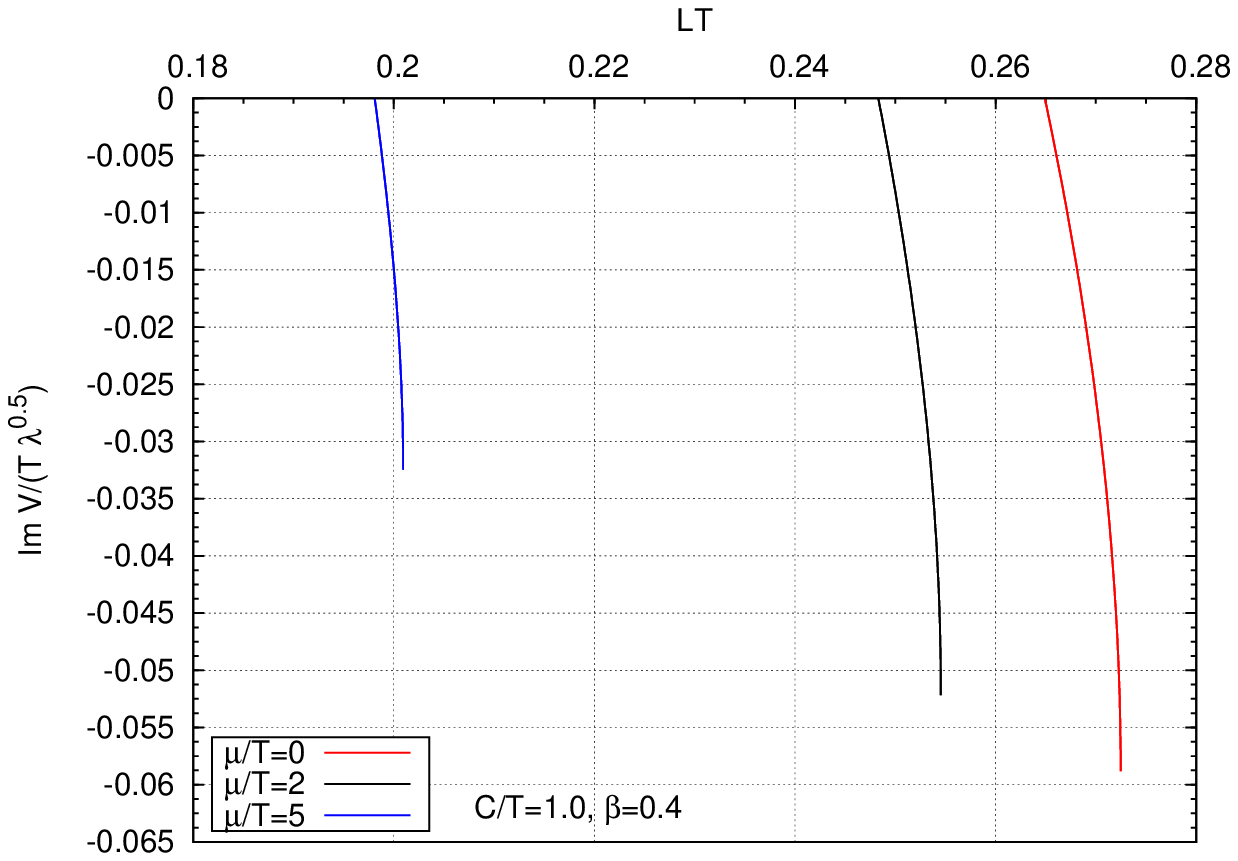}\\
		{\footnotesize parallel case}
	\end{minipage}\hfill
	\begin{minipage}[t]{0.48\textwidth}
		\centering\includegraphics[width=\textwidth]{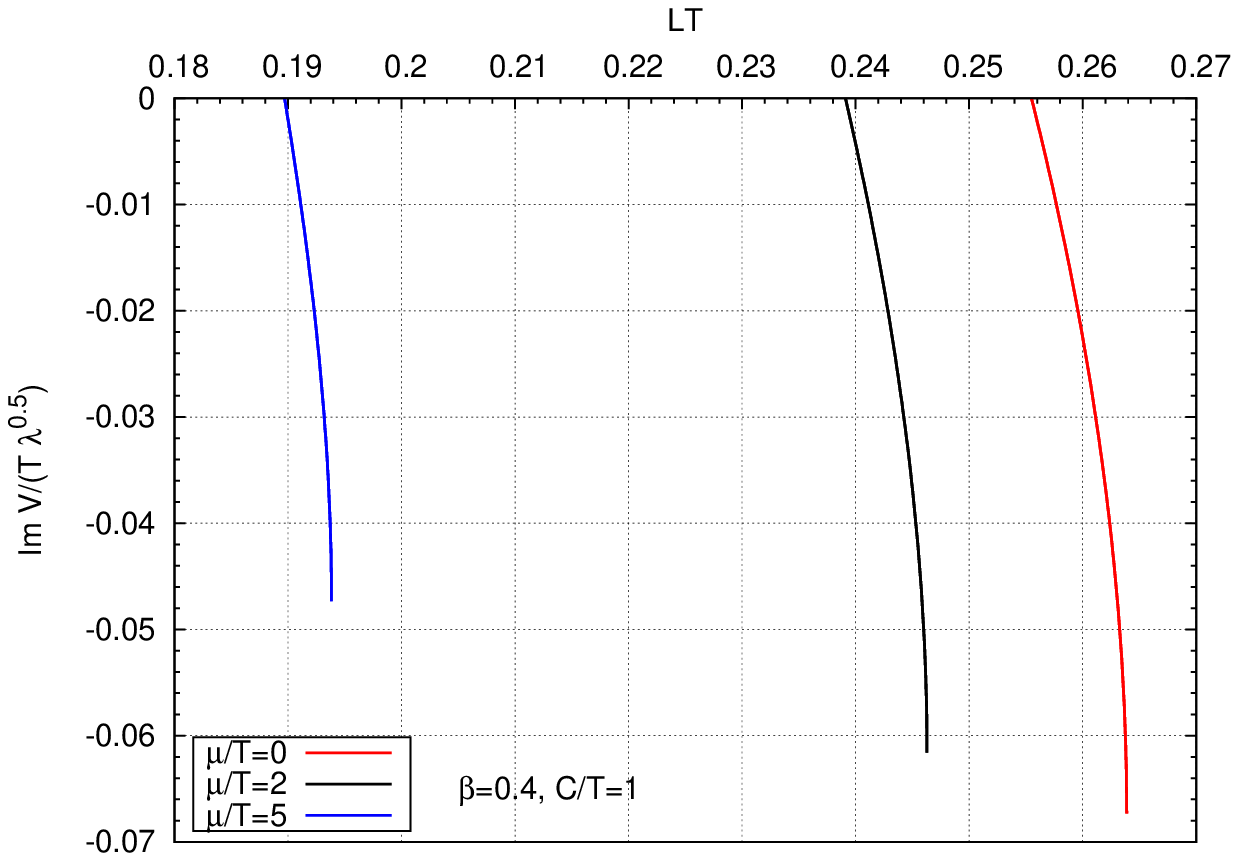}\\
		{\footnotesize perpendicular case}
	\end{minipage}
    \begin{minipage}[t]{0.48\textwidth}
			\centering\includegraphics[width=\textwidth]{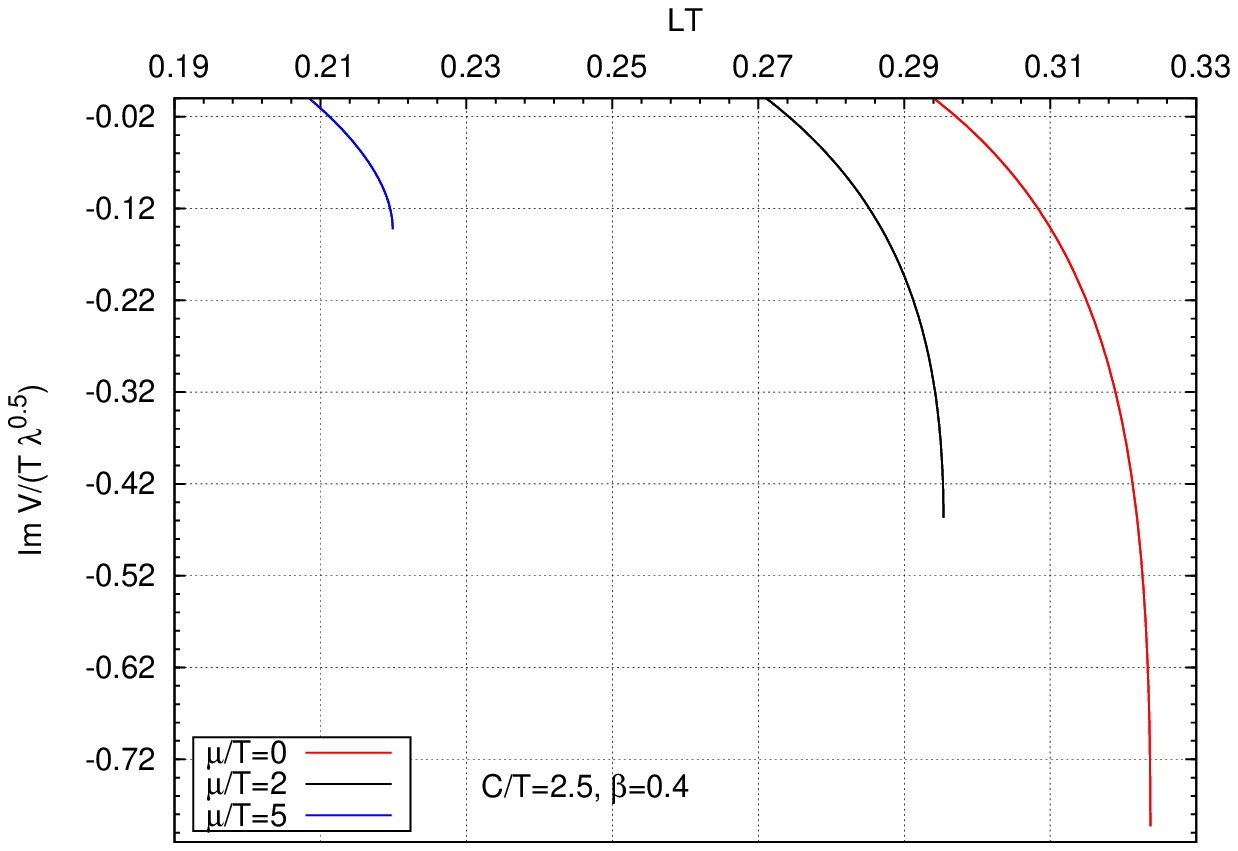}\\
			{\footnotesize parallel case}
		\end{minipage}\hfill
		\begin{minipage}[t]{0.48\textwidth}
			\centering\includegraphics[width=\textwidth]{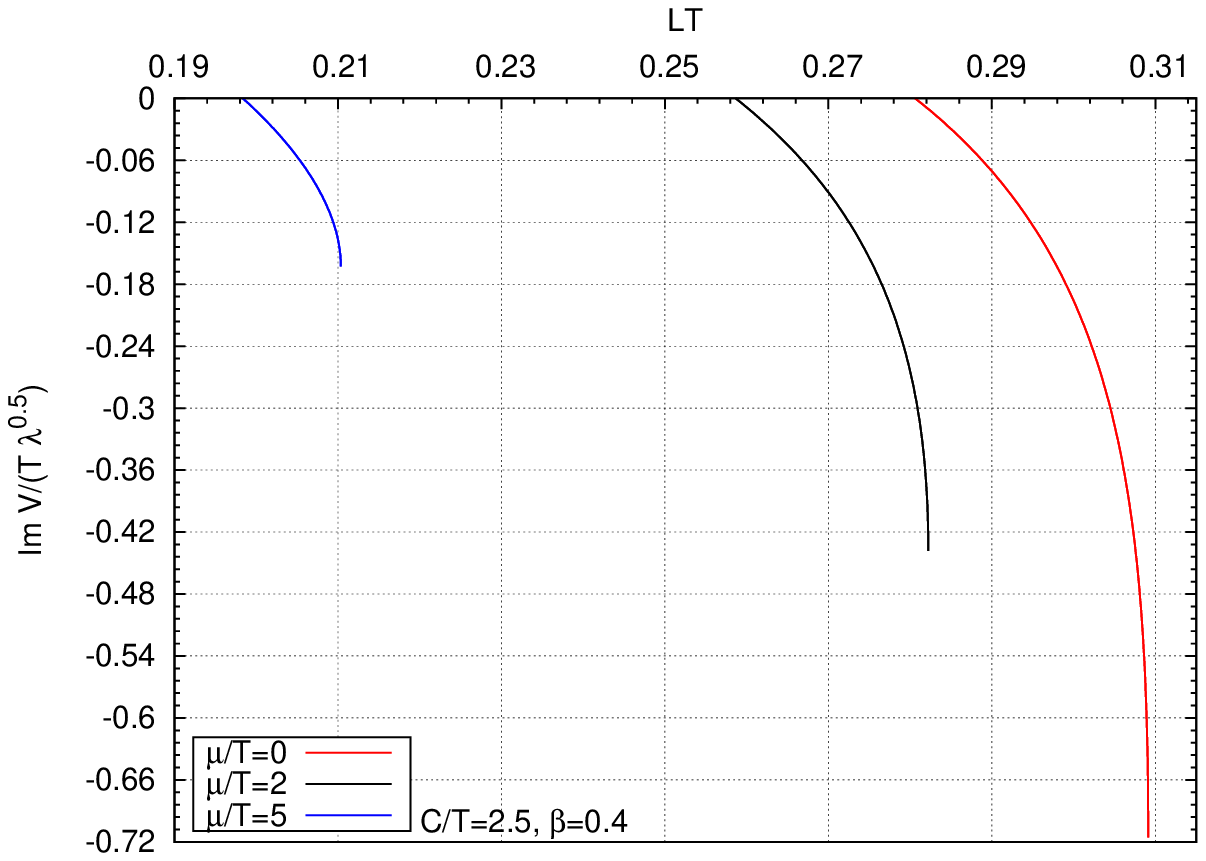}\\
			{\footnotesize perpendicular case}
		\end{minipage}
	\caption{The above plots show the effect of the chemical potential parameter $\frac{\mu}{T}$ and nonconformal parameter $\frac{C}{T}$ on the imaginary part of the $q\bar{q}$ potential.}\label{fig7}
\end{figure}
\noindent In Fig.(\ref{fig7}), we observe the effects of nonconformality and chemical potential on $\mathrm{Im}(V_{q\bar{q}})$. The plots suggest that with the increment in the value of nonconformality parameter $\frac{C}{T}$, the imaginary potential appears at a higher value of $LT$. This directly suggests that nonconformality of the medium increases the distance between $q$ and $\bar{q}$ ($LT$). On the other hand, increment in the value of $\frac{\mu}{T}$ parameter forces the imaginary potential to appear at a smaller value of $LT$ which implies that chemical potential of the medium decreases the distance between $q$ and $\bar{q}$ ($LT$).
It is known that the imaginary part of the $q\bar{q}$ potential is related with the dissociation properties of the quarkonia in the medium. This knowledge suggests that dissociation of the quarkonia becomes easier in the presence of the chemical potential in the medium whereas nonconformality opposes the dissociation process.\\
The computation of the imaginary potential also enables us to comment on the possible thermal widths associated with it. The relationship between these quantities stands to be \cite{im1}-\cite{im7}
\begin{eqnarray}
\Gamma = - \langle\psi\vert \mathrm{Im}(V_{q\bar{q}})\vert\psi\rangle~.
\end{eqnarray}  
It can observed from the computed results of the imaginary potential that the enhancement of the chemical potential $\frac{\mu}{T}$  decreases the imaginary potential which in turn suppresses the thermal width. On the other hand, increase in the value of the confining scale in the theory $\frac{C}{T}$, increases the thermal width. The rapidity $\beta$ creates similar effect as the chemical potential, as it also decreases the thermal width.

\section{Conclusion} \label{sec6}
We now summarize our findings. We have holographically investigated the dynamics of a moving quark-antiquark dipole in a strongly coupled nonconformal plasma with finite quark density. We use the soft-wall dual geometry ($\mathrm{SW}_{T,\mu}$ model) in which a $U(1)$ gauge field is added in the bulk action on the basis of the AdS/CFT dictionary to provide the quark-density vector operator in the boundary field theory. The conformal invariance in the dual field theory is broken by the background dilation which appears as an overall warp factor in the metric. The QGP plasma is moving in a specific direction, namely $x^1$ with a velocity $v<1$. The presence of the Lorentz boost (in order to probe the moving $q\bar{q}$ pair), we have considered two extreme cases of orientation for the $q\bar{q}$ pair, namely, parallel and perpendicular to the direction of boost. In the domain $v<1$, the actions remains real and it leads to time-like Wilson loop.
We then compute the screening length ($LT$) and the real part of the $q\bar{q}$ potential for both orientations by holographically computing the expectation value of the time-like Wilson loop. We observe that in case of the parallel orientation of the $q\bar{q}$ dipole with respect to the direction of boost, a constraint on the turning-point exists which restricts its domain of possible values whereas there is no constraint in case of the transverse case. It is observed that with increasing value of rapidity $\beta$ and chemical potential parameter $\frac{\mu}{T}$, the value of $LT_{max}$ decreases and also the binding energy of $q\bar{q}$ pair (Re$(V_{q\bar{q}})$) reduces. However, the deformation parameter (nonconformality) $\frac{C}{T}$ increases the value of $LT_{max}$ and also increases the value of real part of the $q\bar{q}$ potential. This observation suggests that the the chemical potential and the deformation parameter affects the screening length of the $q\bar{q}$ dipole and Re$(V_{q\bar{q}})$ in an opposite manner. We then take $\beta\rightarrow \infty$ ($v=1$) limit. This makes the action imaginary and the Wilson loop light-like. By holographically computing the light-like Wilson loop we obtain the in-medium energy loss of the moving parton also known as the jet quenching parameter $q_m$. We observe that the presence of the chemical potential and nonconformality both increases the in-medium energy loss of the moving parton. This also suggests that in the high $p_T$ domain, both $\frac{C}{T}$ and $\frac{\mu}{T}$ enhances the gluon radiation of the parton. We compare our results with the observed values of the jet quenching in the RHIC experiments. This comparison suggests that increase in $\mu$ and $C$ lowers the possible allowed domain of temperature for the computed jet quenching parameter to agree with the RHIC observed values ($5\leq q_m \leq 15$ GeV$^2$/fm). We then compute the allowed parameter spaces of $\mu$ and $C$ in order to obtaine the experimentally observed values of $q_m$ at RHIC and LHC at a fixed temperature $T$. Furthermore, we probe the strength of the coupling in our strongly-coupled gauge theory via an order parameter $\mathcal{K}\equiv1.25 \frac{T^3}{q_m}$ and obtain the possible value (minimum) for the coupling constant in order to obtain  $\frac{\eta}{s}\gg\mathcal{K}$.
We then proceed to compute the imaginary part of the $q\bar{q}$ potential by considering the thermal fluctuation of the string world-sheet. It is observed that an increase in the value of rapidity $\beta$ forces the imaginary potential to start from a smaller value of $LT$. However, in case of parallel orientation, imaginary potential vanishes at $\beta=1.2$, whereas at that particular value of $\beta$, $Im(V_{q\bar{q}})\neq0$ for the perpendicular case. The effect of the chemical potential parameter $\frac{\mu}{T}$ is similar to rapidity as increasing $\frac{\mu}{T}$ makes $Im(V_{q\bar{q}})$ to start from a smaller value of $LT$. However, the presence of nonconformality which is incorporated via the deformation parameter $\frac{C}{T}$ forces $Im(V_{q\bar{q}})$ to start from a higher value of $LT$. This suggests that rapidity $\beta$ and chemical potential $\frac{\mu}{T}$ helps in the dissociation process of the quarkonia and nonconformality opposes it. 
To place our findings in proper perspective with the existing results in the literature, we would like to mention that our results are in agreement with earlier findings. For instance, we observe that for a fixed value of the nonconformality parameter, the effect of increasing the  chemical potential with a fixed value of rapidity decreases the values of the real and imaginary potentials but increases the in-medium energy loss.
\\

\section*{Acknowledgements}
A.S. would like to acknowledge the support by Council of Scientific and Industrial Research (CSIR, Govt. of India) for Junior Research Fellowship. S.G. acknowledges the support of the Visiting Associateship programme of IUCAA, Pune. The authors would like to acknowledge the anonymous referees for very useful comments. AS would also like to thank Abhi Mukherjee of University of Kalyani for a fruitful discussion on the computational aspects. 
\section*{Appendix}\label{16}
In this appendix, we write down the functions mentioned in eq.(\ref{17}) by using the string action corresponding to parallel and perpendicular orientation of the $q\bar{q}$ dipole with respect to the direction of boost. 
\subsection{$q\bar{q}$ pair is in tranverse direction with respect to the direction of boost}
Firstly, we use the string action corresponding to the perpendicular orientation of the $q\bar{q}$ pair, given in eq.(\ref{SNG}), and substitute it in eq.(\ref{19}). This leads to the following functions which arise in eq.(\ref{17}).
\begin{eqnarray}
A(\alpha) &=& T^4 \bigg(\frac{t_{\mu}}{\alpha}\bigg)^4 h^2(\alpha)\bigg[f(\alpha)\cosh^2\beta-\sinh^2\beta\bigg]\nonumber\\
A^{\prime}(\alpha) &=&T^3\bigg[ h(\alpha) \bigg[f(\alpha)\cosh^2\beta-\sinh^2\beta\bigg]
\times \bigg[4\bigg(\frac{t_{\mu}}{\alpha}\bigg)^3 h(\alpha)+2\bigg(\frac{t_{\mu}}{\alpha}\bigg)^4 h^{\prime}(\alpha) \bigg] + \bigg(\frac{t_{\mu}}{\alpha}\bigg)^4 h^2(\alpha) f^{\prime}(\alpha)\cosh^2\beta\bigg]\nonumber\\
 B(\alpha)&=&h^2(\alpha)\bigg[\cosh^2\beta-\frac{\sinh^2\beta}{f(\alpha)}\bigg]\nonumber\\
 A^{\prime\prime}(\alpha) &=& T^2\bigg[12 \bigg(\frac{t_{\mu}}{\alpha}\bigg)^2 h^2(\alpha)\bigg[f(\alpha)\cosh^2\beta-\sinh^2\beta\bigg]+ 8 \bigg(\frac{t_{\mu}}{\alpha}\bigg)^3 h(\alpha)h^{\prime}(\alpha) \times\bigg[f(\alpha)\cosh^2\beta-\sinh^2\beta\bigg]\nonumber\\
 && + 4  \bigg(\frac{t_{\mu}}{\alpha}\bigg)^3 h^2(\alpha) f^{\prime}(\alpha)\cosh^2\beta+\bigg[f(\alpha)\cosh^2\beta-\sinh^2\beta\bigg] \times \bigg[8\bigg(\frac{t_{\mu}}{\alpha}\bigg)^3 h(\alpha) h^{\prime}(\alpha)+2\bigg(\frac{t_{\mu}}{\alpha}\bigg)^4{h^{\prime}(\alpha)}^2\nonumber\\
 &&+2\bigg(\frac{t_{\mu}}{\alpha}\bigg)^4 h(\alpha) h^{\prime\prime}(\alpha)\bigg]+f^{\prime}(\alpha)\cosh^2\beta\times\bigg[4h^2(\alpha)\bigg(\frac{t_{\mu}}{\alpha}\bigg)^3+2\bigg(\frac{t_{\mu}}{\alpha}\bigg)^4h(\alpha)h^{\prime}(\alpha)\bigg]\nonumber\\
 &&+\bigg(\frac{t_{\mu}}{\alpha}\bigg)^4 h^2(\alpha) f^{\prime\prime}(\alpha)\cosh^2\beta+2\bigg(\frac{t_{\mu}}{\alpha}\bigg)^4 h(\alpha)h^{\prime}(\alpha) f^{\prime}(\alpha)\cosh^2\beta\bigg]\nonumber
 \end{eqnarray}
 \subsection{$q\bar{q}$ pair is in the same direction with respect to the direction of boost}
Here also, we use the string action corresponding to the perpendicular orientation of the $q\bar{q}$ pair, given in eq.(\ref{ng1}) and substitute it in eq.(\ref{19}). This leads to the following functions which arise in eq.(\ref{17}).
 \begin{eqnarray}
 A(\alpha) &=& T^4 \bigg(\frac{t_{\mu}}{\alpha}\bigg)^4 h^2(\alpha)f(\alpha)\nonumber\\
 A^{\prime}(\alpha) &=&T^3\bigg[4 h^2(\alpha)f(\alpha)\bigg(\frac{t_{\mu}}{\alpha}\bigg)^3 +2\bigg(\frac{t_{\mu}}{\alpha}\bigg)^4 h^{\prime}(\alpha) h(\alpha)f(\alpha) + \bigg(\frac{t_{\mu}}{\alpha}\bigg)^4 h^2(\alpha) f^{\prime}(\alpha)\bigg]\nonumber\\
 A^{\prime\prime}(\alpha) &=& T^2\bigg[12 \bigg(\frac{t_{\mu}}{\alpha}\bigg)^2 h^2(\alpha)f(\alpha)+ 16 \bigg(\frac{t_{\mu}}{\alpha}\bigg)^3 h(\alpha)h^{\prime}(\alpha)f(\alpha)+ 4  \bigg(\frac{t_{\mu}}{\alpha}\bigg)^3 h^2(\alpha) f^{\prime}(\alpha)+2\bigg(\frac{t_{\mu}}{\alpha}\bigg)^4 {h^{\prime}}^2 f(\alpha)\nonumber\\
 &&+2\bigg(\frac{t_{\mu}}{\alpha}\bigg)^4 h(\alpha) h^{\prime\prime}(\alpha) f(\alpha)+4f^{\prime}(\alpha) h^2(\alpha)\bigg(\frac{t_{\mu}}{\alpha}\bigg)^3+2\bigg(\frac{t_{\mu}}{\alpha}\bigg)^4 h(\alpha)h^{\prime}(\alpha)f^{\prime}(\alpha)+\bigg(\frac{t_{\mu}}{\alpha}\bigg)^4 h^2(\alpha) f^{\prime\prime}(\alpha)\nonumber\\
  &&+2\bigg(\frac{t_{\mu}}{\alpha}\bigg)^4 h(\alpha)h^{\prime}(\alpha) f^{\prime}(\alpha)\bigg]\nonumber\\
  B(\alpha)&=&h^2(\alpha)\bigg[\cosh^2\beta-\frac{\sinh^2\beta}{f(\alpha)}\bigg]\nonumber
 \end{eqnarray}
 where
 \begin{eqnarray}
 h(\alpha) &=& \exp\bigg[\bigg(\frac{C}{T}\bigg)^2\bigg(\frac{\alpha}{t_{\mu}}\bigg)^2\bigg];~~~ h^{\prime}(\alpha) =- \frac{2}{T}\bigg(\frac{C}{T}\bigg)^2\bigg(\frac{\alpha}{t_{\mu}}\bigg)^3\exp\bigg[\bigg(\frac{C}{T}\bigg)^2\bigg(\frac{\alpha}{t_{\mu}}\bigg)^2\bigg] \nonumber\\
h^{\prime\prime}(\alpha) &=& \bigg[\frac{4}{T^2}\bigg(\frac{C}{T}\bigg)^4\bigg(\frac{\alpha}{t_{\mu}}\bigg)^6+\frac{6}{T^2}\bigg(\frac{C}{T}\bigg)^2\bigg(\frac{\alpha}{t_{\mu}}\bigg)^4\bigg]\times\exp\bigg[\bigg(\frac{C}{T}\bigg)^2\bigg(\frac{\alpha}{t_{\mu}}\bigg)^2\bigg]\nonumber
\end{eqnarray}
\begin{eqnarray}
f(\alpha) &=& 1- \bigg(1+\frac{1}{3}\bigg(\frac{\mu}{T}\bigg)^2\frac{1}{{t_{\mu}}^2}\bigg)\alpha^4+\bigg(\frac{1}{3}\bigg(\frac{\mu}{T}\bigg)^2\frac{1}{{t_{\mu}}^2}\bigg)\alpha^6\nonumber\\
f^{\prime}(\alpha)&=& \frac{1}{T}\bigg[4\frac{\alpha^5}{t_{\mu}}\bigg(1+\frac{1}{3}\bigg(\frac{\mu}{T}\bigg)^2\frac{1}{{t_{\mu}}^2}\bigg)\alpha^5-2\bigg(\frac{\mu}{T}\bigg)^2\frac{\alpha^7}{{t_{\mu}}^3}\bigg]\nonumber\\
f^{\prime\prime}(\alpha)&=&\frac{1}{T^2}\bigg[-20\bigg(1+\frac{1}{3}\bigg(\frac{\mu}{T}\bigg)^2\frac{1}{{t_{\mu}}^2}\bigg)\frac{\alpha^6}{{t_{\mu}}^2}+14 \bigg(\frac{\mu}{T}\bigg)^2\frac{\alpha^8}{{t_{\mu}}^4}\bigg]\nonumber~.
\end{eqnarray}

\end{document}